\documentclass[preprint,
12pt,showpacs,preprintnumbers,nofootinbib,nobibnotes,amsmath,amssymb,aps,prd]{revtex4-1}
\usepackage{epsfig}
\usepackage{graphicx}
\usepackage{bm}
\usepackage{times}
\usepackage{braket}
\usepackage{color}
\usepackage{slashed}
\usepackage{hyperref}
\usepackage{ulem}
\definecolor{Red}{rgb}{1.,0.,0.}

\definecolor{Blue}{rgb}{0.,0.,1.}

\definecolor{Green}{rgb}{0.,1.,0.}

\definecolor{Gray}{rgb}{0.5,0.5,0.5}

\definecolor{nicered}{rgb}{0.7,0.1,0.1}
\definecolor{nicegreen}{rgb}{0.1,0.5,0.1}
\hypersetup{colorlinks,citecolor=nicegreen,linkcolor=nicered}

\begin{document}

\newcommand{\beq}{\begin{eqnarray}}
\newcommand{\eeq}{\end{eqnarray}}
\newcommand{\ben}{\begin{enumerate}}
\newcommand{\een}{\end{enumerate}}
\newcommand{\non}{\nonumber\\ }
\newcommand{\jpsi}{J/\Psi}
\newcommand{\ppa}{\phi_\pi^{\rm A}}
\newcommand{\ppp}{\phi_\pi^{\rm P}}
\newcommand{\ppt}{\phi_\pi^{\rm T}}
\newcommand{\ov}{ \overline }
\newcommand{\zerot}{ {\textbf 0_{\rm T}} }
\newcommand{\kt}{k_{\rm T} }
\newcommand{\fb}{f_{\rm B} }
\newcommand{\fk}{f_{\rm K} }
\newcommand{\rk}{r_{\rm K} }
\newcommand{\mb}{m_{\rm B} }
\newcommand{\mw}{m_{\rm W} }
\newcommand{\im}{{\rm Im} }
\newcommand{\kks}{K^{(*)}}
\newcommand{\acp}{{\cal A}_{\rm CP}}
\newcommand{\pb}{\phi_{\rm B}}
\newcommand{\xeba}{\bar{x}_2}
\newcommand{\xsba}{\bar{x}_3}
\newcommand{\peas}{\phi^A}
\newcommand{\Dsl}{ D \hspace{-2truemm}/ }
\newcommand{\pvsl}{ p \hspace{-2.0truemm}/_{K^*} }
\newcommand{\esl}{ \epsilon \hspace{-1.8truemm}/ }
\newcommand{\psl}{ p \hspace{-2truemm}/ }
\newcommand{\ksl}{ k \hspace{-2.2truemm}/ }
\newcommand{\lsl}{ l \hspace{-2.2truemm}/ }
\newcommand{\nsl}{ n \hspace{-2.2truemm}/ }
\newcommand{\vsl}{ v \hspace{-2.2truemm}/ }
\newcommand{\zsl}{ z \hspace{-2.2truemm}/ }
\newcommand{\epsl}{\epsilon \hspace{-1.8truemm}/\,  }
\newcommand{\bfkk}{{\bf k} }
\newcommand{\calm}{ {\cal M} }
\newcommand{\calh}{ {\cal H} }
\newcommand{\calo}{ {\cal O} }

\def \appb{{\bf Acta. Phys. Polon. B }  }
\def \cpc{ {\bf Chin. Phys. C } }
\def \ctp{ {\bf Commun. Theor. Phys. } }
\def \epjc{{\bf Eur. Phys. J. C} }
\def \ijmpcs{{\bf Int. J. Mod. Phys. Conf. Ser.} }
\def \jhep{{\bf J. High Energy Phys. } }
\def \jpg{ {\bf J. Phys. G} }
\def \mpla{{\bf Mod. Phys. Lett. A } }
\def \npb{ {\bf Nucl. Phys. B} }
\def \plb{ {\bf Phys. Lett. B} }
\def \ppn{ {\bf Phys. Part. Nucl. } }
\def \ppnp{{\bf Prog.Part. Nucl. Phys.  } }
\def \pr{  {\bf Phys. Rep.} }
\def \prc{ {\bf Phys. Rev. C }}
\def \prd{ {\bf Phys. Rev. D} }
\def \prl{ {\bf Phys. Rev. Lett.}  }
\def \ptp{ {\bf Prog. Theor. Phys. }}
\def \zpc{ {\bf Z. Phys. C}  }
\def \jpg{ {\bf J.Phys.-G-}  }
\def \ap{ {\bf Ann. of Phys}  }


\title{\large{Probing isovector scalar mesons in the charmless three-body $B$ decays}
\vspace{0.4cm}}

\author{Jian Chai$^{1}$}
\author{Shan Cheng$^{1,2}$}\email{scheng@hnu.edu.cn}
\author{Ai-Jun Ma$^{3}$}
\affiliation{$^1$School of Physics and Electronics, Hunan University, Changsha 410082, China \non
$^2$School for Theoretical Physics, Hunan University, Changsha 410082, China \non
$^3$ Department of Mathematics and Physics, Nanjing Institute of Technology, Nanjing 211167, China}

\vspace{1.2cm}

\date{\today}

\vspace{8mm}

\begin{abstract}
We propose to study the multiparticle configurations of isovector scalar mesons, 
saying $a_0(980)$ and $a_0(1450)$, in the charmless three-body $B$ decays by considering the width effects.
Two scenarios of $a_0$ configurations are assumed, 
in which the first one take $a_0(980)$ as the lowest-lying $q{\bar q}$ state and $a_0(1450)$ as the first radial excited state, 
the second one take $a_0(1450)$ as the lowest-lying $q{\bar q}$ state and $a_0(1950)$ as the first radial excited state 
while $a_0(980)$ is not a $q{\bar q}$ state.
Within these two scenarios, we do the PQCD calculation for the quasi-two-body $B \to a_0 \left[\to K{\bar K}/\pi\eta \right] h$ decays 
and extract the corresponding branching fractions of two-body $B \to a_0 h$ decays under the narrow width approximation.
Our predictions show that the first scenario of $a_0(980)$ configuration can not be excluded by the available measurements in $B$ decays,
the contributions from $a_0(1450)$ to the branching fractions in most channels are comparable in the first and second scenarios.
Several channels are suggested for the forthcoming experimental measurements to reveal the multiparticle configurations of $a_0$,
such as the channel $B^0 \to a_0^-(980) \left[\to \pi^-\eta \right] \pi^+$ with the largest predicted branching fraction,
the channels $B^0 \to a_0^{\pm}(1450) \left[\to K^\pm{\bar K}^0, \pi^\pm\eta \right] \pi^\mp$
whose branching fractions obtained in the second scenario is about three times larger in magnitude than that obtained in the first scenario,
and also the channels $B^+ \to a_0^+(1950) \left[ K^+{\bar K}^0/\pi^+\eta \right] K^0$ whose branching fractions
are linear dependent on the partial width $\Gamma_{a_0(1950) \to KK/\pi\eta}$.
\end{abstract}
\maketitle

\section{Introduction}

It is known that the scalar mesons with the masses below and near $1 \, \mathrm{GeV}$,
saying the isoscalar mesons $\sigma/f_0(500)$ and $f_0(980)$, the isovector $a_0(980)$ and the isodoublet $\kappa$, form a $SU(3)$ flavor nonet,
meanwhile, the mesons heavier than $1 \, \mathrm{GeV}$ with including $f_0(1370), f_0(1500)$, $a_0(1450)$ and $K^{\ast}_0(1430)$ make up another nonet.
The underlying assignment of the heavier nonet is almost accepted as the quark-antiquark configuration replenished with some possible gluon content \cite{JaffeIG,CloseZU,AchasovHM,AchasovFH},
while the inner nature of scalar mesons in the lighter nonet is still not clear \cite{WeinsteinGC,WeinsteinGD,Agaev:2017cfz},
even though the compact tetraquark state \cite{Alford:2000mm,Maiani:2004uc,Maiani:2007iw} and the $K\bar{K}$ bound state \cite{Weinstein:1990gu}
are the most favorite two candidates nowadays.
This is easy to understand from the views of spectral analysis at low energy because 
the scalar meson in $q\bar{q}$ configuration has a unit of orbital angular momentum which increases their masses,
in contrast, it is not necessary to introduce the orbital angular momentum
when the scalar meson is being in $q^2\bar{q}^2$ configuration \cite{ChengNB}.
The case becomes different in the weak decays like $B \to f_0(980) l \nu$ with large recoiling,
where the conventional $q\bar{q}$ assignment can be expected to be dominated in the energetic $f_0(980)$
since the possibility to form a tetra-quark state is power suppressed with comparing to the state of quark pair \cite{Cheng:2019tgh},
meanwhile, the final state interaction (FSI) is weak too.
But this argument encounters challenge from the perturbative QCD (PQCD) calculation of $B\to a_0(980) K$ decays
\cite{Shen:2006ms}, where the theoretical predictions of branching fractions are much larger than that of the measured upper limits.
We would like to comment that their calculation is carried out in the static $a_0(980)$ approximation
while the experiment measurement is actually fulfilled by the $\pi\eta$ invariant mass spectral.
It is apparent that the salient property of scalar mesons, say, the large decay width
which cause a strong overlap between resonances and background, and subsequently influence the PQCD prediction.

The width effect of intermediate resonant states have been studied in three-body $B$ decays with a large number of channels
by variable theoretical approaches based on QCD, 
due to the significant physics to understand the hadron structures and also the matter-antimatter asymmetry.
We here highlight some developments in this research filed in the order of different theoretical approaches.
\begin{description}
\item[PQCD]
A global analysis of three-body charmless decays in the type of $B \to V \left[ \to P_1P_2 \right] P_3$\footnote{Here $V,P$ denote the vector and pseudoscalar meson, respectively, and $S$ indicates the scalar meson in the following.
In the fit, only the $P_1P_2 = \pi\pi, \pi K, K\bar K$ channels are considered due to the experiment precision.}
is performed \cite{Li:2021cnd} to determine the lowest several gegenbauer moments of two-meson system,
which are the nonperturbative inputs describing the non-asymptotic QCD correction effect in the light-cone distribution amplitudes (LCDAs).
In Ref.\cite{Rui:2021kbn}, the factorization formulas of PQCD is expanded in the four-body $B$ decay to two $\left[K \pi \right]_{S,P}$ pairs with the invariant mass around the $K^\ast(892)$ resonance,
some further observations like the triple-product asymmetries and the $S$-wave induced direct CP asymmetries are presented
with the interference between different helicity amplitudes. 
Motivated by the measurement of significant derivations from the simple phase-space model in the channels
$B \to K\bar{K} P_1$ and $B_{(s)} \to D_{(s)} P_1P_2$ at B factories and LHC,
the virtual contribution clarified by the experiment collaborations is understood theoretically
by the Breit-Wigner-tail (BWT) effects from the corresponding intermediate resonant states,
say $\rho, \omega$ and $D_{(s)}^\ast$, respectively \cite{Wang:2020nel,Chai:2021kie}.
\item[QCDF]
The QCD factorization (QCDF) formula of amplitudes in three-body $B$ decays \cite{Klein:2017xti} 
is parameterised in a new way where the contributions from valence $u$ and $c$ quark are separated,
and a new source of {\it CP} violation can be generated via the strong phase with the opening of $D\bar{D}$ threshold in the high invariant mass region \cite{Mannel:2020abt}.
Motived by the NNLO $\alpha_s(m_b)$ correction and the finite width effect,
three-body $B$ decay is studied from the point of view of factorisation for the heavy-to-heavy $B \to D \rho \left[ \to \pi\pi \right],
D K^\ast \left[ \to K\pi \right]$ decays
in the kinematics with small invariant mass of dimeson system \cite{Huber:2020pqb}.
Very recently, a novel observation named the forward-backward asymmetry induced $CP$ asymmetry (FBI-{\it CPA})
is introduced in the three-body heavy meson decays, the estimation based on the generalized factorization approach implies that
the FBI-{\it CPA} in the channel $D^\pm \to K^+ K^- \pi^\pm$ is about a milli,
which is at the same order of current experiment measurement capability \cite{Zhang:2021zhr}. In Refs. \cite{Cheng:2020iwk,Cheng:2022vbw}, the finite-width effects of intermediate resonant states in three-body $B/D$ decays is expressed 
by a correlation parameter $\eta_R$ and the evaluation is carried out in QCDF.
\item[LCSRs]
The width effect of intermediate resonant $\rho$ and its radial excited states is discussed in detail in the $P$-wave
$B \to \pi\pi$ transition form factors from the $B$ meson light-cone sum rules (LCSRs) approach \cite{Cheng:2017smj},
revealing the sizeable effects from width and background ($20\% - 30\%$) to the conventional treatment
in the single narrow-width approximation for the LCSRs prediction of the $B \to \rho$ transition form factors.
This result is confirmed by the other independent LCSRs with dipion distribution amplitudes (DAs)
where the hadronic dipion state has a small invariant mass and simultaneously a large recoil \cite{Hambrock:2015aor,Cheng:2017sfk}.
The further studies are carried out for 
the $P$-wave $B \to K\pi$ form factors with the isodouble intermediate resonances
$K_0^\ast$ \cite{Descotes-Genon:2019bud} and the $B_s \to K\bar K$ form factors
with the isoscalar scalar intermediate resonances $f_0(980)$ and $f_0(1450)$ \cite{Cheng:2019tgh}.
\end{description}

The above considerations mainly focus on the $P$-wave and isoscalar $S$-wave contributions from the intermediate resonant states, 
while the study of isovector scalar intermediate resonance is still missing.
In this paper we will demonstrate this issue in the framework of PQCD approach.
The motivations of this study is twofold.
Firstly, we perform the PQCD prediction of $B \to a_0(980) \left[ \to \eta \pi \right] K$ decays go beyond the single pole approximation, 
trying to explain the measurement status.
Secondly, we consider the roles of $a_0(1450)$ and $a_0(1950)$ in the $B \to {\bar K}K K$ decays
inspired by the recent measurements of charm meson decays where $a_0(1450)$ and $a_0(1950)$ are observed
in the $K\bar K$ invariant mass spectral \cite{CLEO:2008msk,LHCb:2015lnk,BaBar:2015kii},
supplementing to the $B \to \eta \pi K$ decays observed firstly
at Crystal Barrel Collaboration long time ago \cite{CrystalBarrel:1994arw,CrystalBarrel:1995dzq}.
The study would be executed in parallel by taking two different scenarios of $a_0$ states,
where the first one says that $a_0(980)$ is the lowest lying $q\bar{q}$ state and $a_0(1450)$ is the first excited state,
and the second one states that $a_0(1450)$ and $a_0(1950)$ are the lowest lying $q\bar{q}$ state and the first excited state, respectively. 
Our calculations show that the $q{\bar q}$ configuration of $a_0(980)$ is not be excluded by the available measurements in $B$ decays, 
which confirms the statements we made above.
Predictions in this work would help us to probe the inner structure of energetic isovector scalar mesons. 
For examples, 
(a) the channel $B^0 \to a_0^-(980) \left[\to \pi^-\eta \right] \pi^+$ has the largest branching fraction under the $q{\bar q}$ configuration of $a_0(980)$,
(b) the branching fractions of channels $B^0 \to a_0^{\pm}(1450) \left[\to K^\pm{\bar K}^0, \pi^\pm\eta \right] \pi^\mp$
obtained in the second scenario is about three times larger in magnitude than that obtained in the first scenario, 
even though the result obtained from two scenarios are close to each other in the most channels with the intermediate state $a_0(1450)$,  
(c) the branching fractions of channels $B^+ \to a_0^+(1950) \left[ K^+{\bar K}^0/\pi^+\eta \right] K^0$ 
are linear dependent on the partial width $\Gamma_{a_0(1950) \to KK/\pi\eta}$ in the second scenario.

This article is organized as follows.
In section \ref{sec-framework}, the framework of PQCD approach to deal with the resonance contribution in three-body $B$ decays
is briefly described in turns of kinematics and dynamics.
Section \ref{sec-numerics} presents the PQCD predictions of the $B \to a_0 \left[ \to K\bar K, \eta \pi \right] h$ decays with some discussions.
We summary in section \ref{sec-conclusion}. 
The PQCD predictions on $B_s$ decays are presented in appendix \ref{sec-Bs}, 
and the factorization formulas of the related quasi-two-body decay amplitudes are listed in appendix \ref{sec-appx-amplitudes}.

\section{Kinematics and Dynamics}\label{sec-framework}

Concerning three-body $B$ decays,
there are three typical kinematical configurations in the physical Dalitz plot of two independent invariant mass by considering the four-momentum conservation,
in which only the kinematical region with collinear decay products can be calculated reliably
from the perturbative theory based on the factorization hypothesis \cite{Chai:2021kie}.
The other two kinematical regions with the three energetic decay products and a soft decay product configurations
are either in lack of the rigorous factorization proof or beyond the available perturbative picture of heavy meson decays.
Collinear decay products means that two energetic hadrons move ahead with collinear momenta 
while the rest one recoiling back\footnote{$E_i \sim m_B/2$
and $E_j + E_k \sim m_B/2$ in the massless approximation of final mesons.},
corresponding to the intermediate parts of three edges in the Dalitz plot.

The matrix element from vacuum to collinear two meson system sandwiched with certain two quark operators is defined by the dimeson DAs,
the chirally even two quark dimeson DA is quoted for example as \cite{Polyakov:1998ze}
\beq
\langle M_1^a(p_1) M_2^b(p_2) \vert \, \bar{q}_f(xn) \, \tau \, q_{f^\prime}(0) \, \vert 0 \rangle
= \kappa_{ab} \int dz \, e^{i zx (p_R \cdot n)} \, \Phi_{M_1M_2}^{ab, f f^\prime} (z, \zeta, s) \,,
\label{eq:dimeson-DA}
\eeq
where the indexes $f, f^\prime$ respect the (anti-)quark flavor; $a, b$ indicate the electric charge of each meson, 
$\kappa_{ab}$ is the isospin symmetry coefficient which in the case of dipion system reads $\kappa_{+-/00} = 1$ and $\kappa_{+0} = \sqrt{2}$, 
$p_R = k_1 + k_2$ is the total momentum of dimeson state, 
$\tau = 1/2, \tau_3/2$ correspond to the isoscalar and isovector dimeson DAs, respectively.
The generalized dimeson DA $\Phi_\parallel^{ab, f f^\prime}$ is characterised by three independent kinematical variables,
saying the momentum fraction $z$ carried by the antiquark, the longitudinal momentum fraction carried by one of the mesons $\zeta = p_1^+/p_R^+$
and the invariant mass squared $s = p_R^2$.
Besides the conventional Gegenbauer expansion stemmed from the eigenfunction of QCD evolution equation,
the partial wave expansion considered in the dimeson system contributes the other Legendre polynomial $C_l^{1/2}$.
The double expansion of two quark dimeson DA is written as
\beq
\Phi_{M_1M_2}^{I=1} (z, \zeta, s, \mu) = 6 z (1-z) \sum_{n=0, {\rm even}}^{\infty} \, \sum_{l=1, {\rm odd}}^{n+1}
B_{nl}^{I=1}(s, \mu) \, C_n^{3/2}(2z-1) \, C_l^{1/2}(2\zeta-1) \,,
\label{eq:dimeson-da-Expansion}
\eeq
here the even Gegenbauer index $n$ and the odd partial-wave index $l$ are guaranteed by the $C$ parity.
For the expansion coefficients $B_{nl}$, they have the similar scale dependence as the Gegenbauer moments of single pion and rho mesons.
In the narrow width approximation in the vicinity of the resonance, dimenson DAs reduce to the DAs of the relative resonance,
indicating that the Gegenbuer moments of the intermediate resonance is actually proportional to
the expansion coefficient at zero point with the lowest partial wave, says $a_n^R(\mu) \propto B_{n1}(s=0, \mu)$.
In this way, the decay constant of intermediate resonance is proportional to the product
of its decay width with the imaginary part of first expansion coefficient at the resonant pole,
that is $f_R \propto \Gamma_R \, {\rm Im}[B_{01}(m_R^2)]$ \cite{Cheng:2019hpq}.

With this definition, the dimeson DAs are the most general objects to describe the dimeson mass spectrum in hard production processes whose
asymptotic formula indicates the information of the deviation from the unstable intermediate resonant meson DAs.
After integrating over the momentum fraction of antiquark, the isovector scalar dimeson DA in our interest is normalised to timelike meson form factor as
\beq
\int_0^1 dz  \, \Phi_{M_1M_2}^{I=1} (z, \zeta, s) = \left( 2 \zeta - 1\right) {\it \Gamma}_{M_1M_2}^{I=1}(s) \,,
\label{eq:dimeson-norm}
\eeq
where the timelike form factor at zero energy point is normalised to unit as ${\it \Gamma}_{M_1M_2}^{I=1}(0) = 1$.
When the invariant mass of dimeson system is small, 
the higher $\mathcal{O}(s)$ terms in the expansion of coefficient $B_{nl}(s, \mu)$ around the resonance pole can be safely neglected
due to the large suppression $\mathcal{O}(s/m_b^2)$ in contrast to the energetic dimeson system in $B$ decay,
so the relation $B_{n1}(s, \mu)  \to a_n(\mu) \, {\it \Gamma}_{M_1M_2}^{I=1}(s)$ can be obtained in the lowest partial wave approximation.
This argument induces the basic assumption in PQCD that the energetic dimeson DAs can be deduced from the DAs of resonant meson
by replacing the decay constant by the timelike form factor.

The isovector scalar form factor of $K\bar K$ and $\pi \eta$ systems are defined by the local matrix elements sandwiched by two quark operator
\cite{Donoghue:1990xh,Albaladejo:2015aca}
\beq
\langle K^- K^0(\pi^- \eta) \vert \bar u(0) \frac{\tau_3}{2} d(0) \vert 0 \rangle
= \frac{m_{\pi}^2}{m_u+m_d} {\it \Gamma}^{I=1}_{K\bar K(\pi \eta)}(s) \equiv B_0 {\it \Gamma}^{I=1}_{K\bar K(\pi \eta)}(s) \,
\label{eq:isovector-ff-definition}
\eeq
with the normalization conditions ${\it \Gamma}^{I=1}_{K\bar K}(0) = 1$ and ${\it \Gamma}^{I=1}_{\pi \eta}(0) = \sqrt{6}/3$.
In the single resonance approximation, we insert a $a_0$ state in the matrix elements
\beq
\langle K^- K^0 (\pi^+ \eta) \vert \bar u(0) \frac{\tau_3}{2} d(0) \vert 0 \rangle
&\approx& \frac{\langle K^- K^0 (\pi^+ \eta) \vert a_0^- \rangle \langle a_0^- \vert \bar u(0) \frac{\tau_3}{2} d(0) \vert 0 \rangle}{\mathcal{D}_{a_0}} \non
&=& \frac{g_{a_0 K\bar K(\pi \eta)} m_{a_0} \bar f_{a_0}}{\mathcal{D}_{a_0}} \,,
\label{eq:a0-insertion}
\eeq
and ultimately arrive at
\beq
{\it \Gamma}^{I=1}_{K\bar K(\pi \eta)}(s) = \frac{g_{a_0 K\bar K(\pi \eta)} m_{a_0} \bar{f}_{a_0}}{B_0 \mathcal{D}_{a_0}} \,.
\label{eq:isovector-ff}
\eeq
Several comments are supplemented in turns to demonstrate this expression.
\begin{itemize}
\item
The decay constants of scalar meson are defined with the scalar and vector currents,
\beq
&&\langle S \vert \bar u(0) \frac{\tau_3}{2} d(0) \vert 0 \rangle = m_S \bar f_S \,, \non
&&\langle S(p) \vert \bar u(0) \gamma_\mu \frac{\tau_3}{2} d(0) \vert 0 \rangle = p_\mu f_S \,.
\eeq
They are related by the equations of motion $\frac{m_S f_S}{m_u-m_d} = \bar f_{S}(\mu)$,
indicating that the neutral scalar meson can not be produced via the vector current
because of the charge conjugation invariance or the conservation of vector current, but the constant $\bar f_S$ is still finite.
\item
Under the narrow $a_0$ approximation, the matrix element of strong decay is defined by the coupling {\cite{Wang:2020saq}}
\beq
\langle K^- K^0 (\pi^+ \eta) \vert a_0^- \rangle
= g_{a_0 K\bar K(\pi \eta)}
= \sqrt{\frac{8\pi m^2_{a_0} \Gamma_{{a_0}\to K\bar K(\pi \eta)}}{q_0}} \,
\label{eq:ga0KK-coup}
\eeq
with the energy independent partial decay width\footnote{The partial widths of
$a_0^0 \to K\bar K$ decays have the relations $\Gamma_{a_0^0\to K^+ K^-}=\Gamma_{a_0^0\to K^0 \bar K^0} = \Gamma_{a_0\to K\bar K}/2$.}
$\Gamma_{{a_0}\to K\bar K(\pi \eta)}$.
In the definition, $q_0 = q(m^2_{a_0})$ is the magnitude of daughter meson ($K(\pi)$ or $\bar K(\eta)$) momentum  
\beq
q(s) = \frac{1}{2}\sqrt{\left[s-(m_{K(\pi)}+m_{\bar K (\eta)})^2\right]\left[s-(m_{K(\pi)}-m_{\bar K (\eta)})^2\right]/s} \,
\label{eq:daughter-meson-momentum}
\eeq
at $a_0$ mass. We take the renormalized mass of $a_0$ rather than the pole mass obtained from $T$-matrix analysis, 
since the mass and width parameter are strongly distorted with lying just below the opening of $K{\bar K}$ channel 
and hence generating an important cusp-like behaviour in the resonant amplitude \cite{Abele:1998qd}.
Actually, $q = \sqrt{s} \beta(s)$ with $\beta(s)$ being the nondimensional phase space factor of $K\bar K(\pi \eta)$ system,
which reflects the information of momentum difference described by the variable $\zeta$ mentioned in the dimeson DAs.
\item
We take the conventional energy-dependent Breit-Wigner denominator for $a_0^\prime$ and $a_0^{\prime \prime}$ mesons\footnote{Hereafter
we use the abbreviations $a_0, a_0^\prime$ and $a_0^{\prime\prime}$ to denote $a_0(980), a_0(1450)$ and $a_0(1950)$, respectively.},
\beq
\mathcal{D}_{a_0^\prime} = m_{a_0^\prime}^2 - s - i m_{a_0^\prime} \, \Gamma^{\rm tot}_{a_0^\prime} \, \frac{q(s)}{q_0}\frac{m_{a_0^\prime}}{\sqrt{s}} \,,
\label{eq:BW-energy}
\eeq
where $\Gamma^{\rm tot}_{a_0^\prime}$ is the total decay widths of resonant state meson $a_0^\prime$.
For the meson $a_0(980)$, we consider the Flatt${\rm \acute{e}}$ model \cite{Flatte:1976xu} 
\beq
&&\mathcal{D}_{a_0} = m_{a_0}^2-s-i (g_{\pi \eta}^2 \beta_{\pi \eta} + g_{K{\bar K}}^2 \beta_{K{\bar K}}) \,,
\label{eq:Flatte}
\eeq
the coupling constants $g_{\pi \eta}=0.324$ GeV and $g_{K{\bar K}}^2/g_{\pi \eta}^2=1.03$ is fixed by the isobar model fits \cite{Abele:1998qd}. 
Furthermore, we can get $g_{a_0 \pi\eta}=2.297$ GeV and $g_{a_0 K{\bar K}}=2.331$ GeV with the relations
$g_{K{\bar K}}=g_{a_0 K{\bar K}}/(4\sqrt{\pi})$ and $g_{\pi \eta}=g_{a_0 \pi\eta}/(4\sqrt{\pi})$. We mark that, in the $a_0 \to \pi\eta$ channel, the phase factor $\beta_{K{\bar K}}$ could also be pure imaginary number 
when the invariant mass of $\pi \eta$ state is small than the threshold value of $K{\bar K}$ state, 
the contribution from this region interacts destructively with that from the rest region of $\pi \eta$ invariant mass. 
\end{itemize}

With rearranging the kinematical variable $\zeta$ into the daughter meson momentum $q(s)$ and considering the $SU(3)$ symmetry,
the matrix element from vacuum to S-wave $K\bar K/\pi\eta$ state can be decomposed as ~\cite{ChengNB}
\beq
\Phi_{K\bar K (\pi \eta)}(z,s)=\frac{1}{\sqrt{2N_c}}\left[{\psl_R}\phi(z,s)+\sqrt s\phi^{s}(z,s)+\sqrt s(\vsl\nsl-1) \phi^{t}(z,s) \right] \,.
\label{def-wavefun-Kpi}
\eeq
In the lowest partial-wave accuracy, the twist 2 LCDA is written as \cite{Cheng:2013fba}
\beq
\phi(z,s)=\frac{{\it \Gamma}_{K\bar K(\pi \eta)}(s)}{2\sqrt{2N_c}}
6z(1-z)\left[\frac{f_S}{\bar f_S(\mu)} +\sum^{\infty}_{m=1} B_{m}(\mu) \, C^{3/2}_m(2z-1) \right] \,,
\label{def-wavefun-twist2}
\eeq
with $B_0(\mu) \equiv f_S/\bar f_S(\mu) \gg 1$.
It is clear that the even Gegenbauer coefficients $B_m$ are suppressed and the odd Gegenabauer moments
is dominated in the twist 2 LCDA of scalar meson, this is definitely different from the $\pi$ and $\rho$ mesons in which the odd moments vanish.
The twist 3 LCDAs are
\beq
&&\phi^{s}(z,s)=\frac{{\it \Gamma}_{K\bar K(\pi \eta)}(s) }{2\sqrt{2N_c}} \left[1+\sum^{\infty}_{m=1} a_{m}(\mu) \, C^{1/2}_m(2z-1) \right] \,, \non
&&\phi^{t}(z,s)=\frac{{\it \Gamma}_{K\bar K(\pi \eta)}(s) }{2\sqrt{2N_c}}(1-2z) \left[1+\sum^{\infty}_{m=1} b_{m}(\mu) \, C^{1/2}_m(2z-1) \right] \,.
\label{def-wavefun-twist3}
\eeq
The definitions of $B$ meson and light meson wave functions and the models of their LCDAs,
as well as the basic procedures of PQCD approach to deal with the so called quasi two-body $B$ decays as a marriage problem,
can be found in detail in Ref. \cite{Chai:2021kie}.

\begin{figure}[t]
\vspace{-2mm}
\begin{center}
\includegraphics[width=1.0\textwidth]{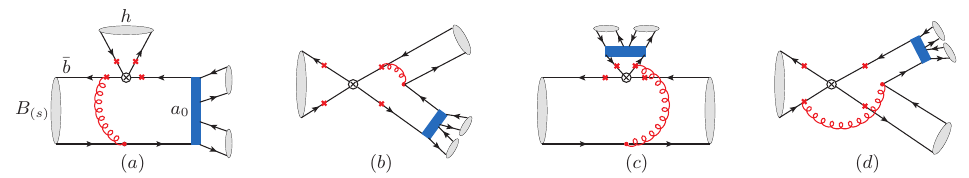}
\vspace{-8mm}
\caption{Typical feynman diagrams for the $B\to a_0 \left[\to K\bar K/\pi \eta\right] h$ decays.}
\label{fig-feyndiag}
\end{center}
\end{figure}

In figure \ref{fig-feyndiag}, we depict the typical feynman diagrams of the $B\to a_0 \left[ \to K\bar K/\pi \eta \right] h$ decays with $h=\pi, K$
in the PQCD approach, in which the symbols $\otimes$ and $\times$ denotes the vertex of weak interaction and the possible attachments of hard gluons, respectively, the rectangle indicates the intermediate resonant states $a_0$ and the subsequent strong decays $a_0 \to K\bar K/\pi\eta$.
In the $B$ meson rest frame, the explicit definitions of kinematics in the
$B(p_B) \to R(p_R) \left[ \to h_1(p_1) h_2(p_2) \right] \,h_3(p_3)$ decays are considered as follow,
\beq
&&p_B = \frac{m_B}{\sqrt2}(1,1,{\bf 0}) \,, \quad \quad \quad \;\; k_B = \left(0, \frac{m_B}{\sqrt2} x_B, {\bf k_{B T}} \right) \,, \non
&&p_R = \frac{m_B}{\sqrt2}(1, \xi, {\bf 0}) \,, \quad \quad \quad \;\; k_R = \left(\frac{m_B}{\sqrt2} z, 0, {\bf k_{T}} \right) \,, \quad \non
&&p_3 = \frac{m_B}{\sqrt2}(0,1-\xi,{\bf 0}) \,, \quad \quad k_3=\left(0, \frac{m_B}{\sqrt2}(1-\xi)x_3, {\bf k_{3T}} \right) \,,
\label{kinematics}
\eeq
where $k_B, k_R$ and $k_3$ are the momenta carried by the antiquark in the meson states with the momentum fractions $x_B, z$ and $x_3$, respectively.
The new variable $\xi \equiv s/m^2_B$ indicates the momentum transfer from $B$ meson to resonant state $R$.
The differential branching ratios for the quasi-two-body $B_{(s)}\to a_0 \left[ \to K\bar K/\pi\eta \right] h$ decays is written as \cite{PDG-2020}
\beq
\frac{d{\mathcal B}}{d\zeta} = \frac{\tau_B \, q_h(s) \, q(s)}{64 \, \pi^3 \, m_{B_{(s)}}} \, \overline{\vert {\mathcal A} \vert^2} \,,
\label{eq:diff-bra}
\eeq
in which daughter meson momentum $q(s)$ has been defined in Eq. (\ref{eq:daughter-meson-momentum}),
and $q_{h}(s)$ is the magnitude of momentum for the bachelor meson $h$
\beq
q_{h}(s)=\frac{1}{2}\sqrt{\big[\left(m^2_{B}-m_{h}^2\right)^2 -2\left(m^2_{B}+m_{h}^2\right)s+s^2\big]/s} \,.
\eeq
The decaying amplitudes is exactly written as a convolution of the hard kernel $H$ with the hadron distribution amplitudes (DAs)
$\phi_B, \phi_h$ and $\phi_{K\bar K, \pi\eta}$
\beq
&~&\mathcal{A} \left( B_{(s)}\to a_0 \left[\to K\bar K/\pi\eta\right] h \right) 
\equiv  \big\langle \left[ K\bar K/\pi\eta \right]_{a_0} h \big\vert \, \mathcal{H}_{eff} \, \big\vert B_{(s)} \big\rangle \non
&=& \phi_B(x_1,b_1, \mu) \otimes H(x_i,b_i,\mu) \otimes \phi_{K\bar K/\pi\eta}(x, b, \mu) \otimes \phi_{h}(x_3, b_3, \mu) \,,
\label{eq:quasi-2body-pQCD}
\eeq
in which $\left[ K\bar K/\pi\eta \right]_{a_0}$ indicates the dimeson system in our interesting,
$\mu$ is the factorization scale, $b_i$ are the conjugate distances of transversal momenta.
We present the expressions of amplitudes ${\mathcal A}$ for the considered decaying processes in the appendix \ref{sec-appx-amplitudes}. 
Under the narrow width approximation 
\beq
\mathcal{A} &=& \int ds \, \frac{ \big\langle K{\bar K}/\pi\eta \big\vert a_0 \big\rangle \big\langle a_0 h \big\vert \, \mathcal{H}_{eff} \, \big\vert B_{(s)} \big\rangle} {[m_{a_0}^2 - s - im_{a_0} \Gamma_{a_0}(s)]} 
\rightarrow \big\langle K{\bar K}/\pi\eta \big\vert a_0 \big\rangle \big\langle a_0 h \big\vert \, \mathcal{H}_{eff} \, \big\vert B_{(s)} \big\rangle \,,
\label{eq:A_NA}
\eeq
we can extract the branching fractions of two-body decays from the quasi-two-body decays by 
\beq
{\mathcal B}\left( B_{(s)}\to a_0 \left[\to K\bar K/\pi\eta\right] h \right) \approx {\mathcal B}\left( B_{(s)}\to a_0 h \right) \cdot 
{\mathcal B}\left( a_0 \to K\bar K/\pi\eta \right) \,.
\eeq

\section{Numerics and Discussions}\label{sec-numerics}

\begin{table}[b]
\begin{center}
\caption{Inputs of the single mesons (in units of GeV) and the Wolfenstein parameters \cite{PDG-2020}.}
\vspace{2mm}
\label{tab1-paras}
\begin{tabular}{l}
\hline\hline
$m_{B^{0}}=5.280 \quad\; m_{B^{\pm}}=5.279 \quad\; m_{B^0_s}=5.367 \quad\; f_{B} = 0.190 \quad\;\;\;\; f_{B_s} = 0.230 \;$\\
$m_{\pi^\pm}=0.140 \quad\; m_{\pi^0}=0.135 \quad\;\, m_{K^\pm}=0.494 \quad\,  m_{K^0}=0.498 \quad\; m_{\eta}=0.548 \;$ \\
$f_{\pi^\pm}=0.130 \quad\;\; f_{\pi^0}=0.156 \quad \;\;\; m_{a_0}=0.980  \quad \;\;  m_{a_0^\prime}=1.474 \quad
\;\; m_{a_0^{\prime\prime}}=1.931 \;$ \\
$\Gamma_{a_0}=0.075 \pm 0.025 \quad\quad\;\; \Gamma_{a_0^\prime} = 0.265 \pm 0.013 \quad\quad\;\;
\Gamma_{a_0^{\prime\prime}}=0.271 \pm 0.036$ \cite{BaBar:2015kii} \; \\
$\lambda=0.22650\pm 0.00048  \;\;\;\,  A=0.790^{+0.017}_{-0.012}  \;\;\;\,  
\bar{\rho} = 0.141^{+0.016}_{-0.017} \;\;\;\, \bar{\eta}= 0.357\pm 0.011 $\\
\hline\hline
\end{tabular}
\end{center}
\end{table}

In table \ref{tab1-paras}, we present the PDG averaged value for the masses and total widths of single mesons,
as well as the Wolfenstein parameters of CKM matrix.
$B_{(s)}$ meson wave function relies on the three independent parameters,
saying the mass $m_B$, the decay constant $f_B$ and the first inverse moment $\omega_B$.
For the inverse moment $\omega_B$, we take the interval $\omega_B (1 \, {\rm GeV}) = 0.40 \pm 0.04 \, {\rm GeV}$
and $\omega_{B_s} (1 \, {\rm GeV}) = 0.50 \pm 0.05 \, {\rm GeV}$ obtained by the QCD sum rules \cite{Braun:2003wx} 
with considering smaller uncertainty. 
The mean lifes of $B$ mesons are also taken from PDG,
they are $\tau_{B^\pm} = 1.638 \times 10^{-12} \, {\rm s}$, $\tau_{B^0} = 1.520 \times 10^{-12} \, {\rm s}$ 
and $\tau_{B_s} = 1.509 \times 10^{-12} \, {\rm s}$.

The PDG value of light meson decay constant follows from the lattice QCD average $f_{K^+}/f_{\pi^+} = 1.193$ \cite{Aoki:2016frl}.
We truncate to the second moments for the Gegenbauer expansion of leading twist LCDAs,
and take $a^\pi_1 = 0$ and $a^\pi_2 (1 \, {\rm GeV}) = 0.270 \pm 0.047$
obtained recently from the LCSR fit \cite{Cheng:2020vwr} of the pion electromagnetic form factor\footnote{This result
agrees with the previous LCSRs extractions from spacelike pion electromagnetic form factor \cite{Agaev:2005gu},
$B \to \pi$ form factor \cite{Ball:2005tb,Duplancic:2008ix,Khodjamirian:2011ub}, and also the QCD sum rule prediction \cite{Ball:2006wn},
but much larger than the recent lattice QCD evaluation ($a^\pi_2 (1 \, {\rm GeV}) = 0.130$)
with the new developed momentum smearing technique \cite{RQCD:2019osh}.}.
For the kaon meson, we take the lattice result obtained by using $N_f=2+1$ sea quarks and the domain-wall fermions \cite{Arthur:2010xf},
say, $a^K_1 (1 \, {\rm GeV}) = 0.060 \pm 0.004$ and $a^K_2 (1 \, {\rm GeV}) = 0.175 \pm 0.065$,
which is comparable with the QCD sum rules calculations \cite{Khodjamirian:2004ga,Ball:2006wn}
and the result from Dyson-Schwinger equations with dynamical chiral spontaneously breaking (DCSB)-improved kernel \cite{Shi:2015esa}.
We takes the chiral masses at $m_0^\pi = 1.913 \, {\rm GeV}, m_0^K = 1.892 \,{\rm GeV}$
with considering the well-known chiral perturbative theory ($\chi{\rm PT}$) relations \cite{Leutwyler:1996qg}
\beq
m_0^\pi = \frac{m_\pi^2 {\cal R}}{2 m_s} \,, \quad\quad\quad
m_0^K = \frac{m_K^2}{m_s [1 +  \frac{1}{{\cal R}} (1 - \frac{{\cal R}^2-1}{4 {\cal Q}^2}) ]}  \,,
\eeq
in which ${\cal R} \equiv 2m_s/(m_u+m_d) = 24.4 \pm 1.5$,
${\cal Q}^2 \equiv [m_s^2 - (m_u + m_d)^2/4]/(m_d^2 - m_u^2) = (22.7 \pm 0.8)^2$,
the current quark masses are $\overline{m}_s (1 \, {\rm GeV}) = 0.125 \,{\rm GeV}$,
$\overline{m}_d(1 \, {\rm GeV}) = 0.0065  \,{\rm GeV}$ and $\overline{m}_u (1 \, {\rm GeV}) = 0.0035 \,{\rm GeV}$.
For the twist 3 LCDA, we only take into account the asymptotic terms in the numerical analysis.

Concerning the intermediate resonant isovector scalar states $a_0$s,
the main inputs are the timelike form factor entered in each LCDA and the Gegenbauer moments in the leading twist LCDA.
To reveal the timelike form factor described in Eq. (\ref{eq:isovector-ff}),
we use the QCD sum rules predictions on the decay constants \cite{ChengNB},
they are $\bar f_{a_0}(1 \, {\rm GeV}) = 0.365 \pm 0.020 \, {\rm GeV}$ and $\bar f_{a_0^\prime}(1 \, {\rm GeV}) = - 0.280 \pm 0.035 \, {\rm GeV}$
obtained in the first scenario where $a_0$ is treated as the lowest lying $q\bar{q}$ state and $a_0^\prime$ as the first excited state,
and $\bar f_{a_0^\prime}(1 \, {\rm GeV}) = 0.460 \pm 0.050 \, {\rm GeV}$ and $\bar f_{a_0^{\prime\prime}}(1 \, {\rm GeV}) = 0.390 \pm 0.040 \, {\rm GeV}$
obtained in the second scenario where $a_0^\prime$ is the lowest lying $q\bar{q}$ state and $a_0^{\prime\prime}$ as the first excited state.
As shown in Eq. (\ref{eq:ga0KK-coup}), the strong coupling constants $g_{a_0K\bar K}$ and $g_{a_0\pi\eta}$ are decided by the partial decay width,
which are fixed by the following considerations
\begin{itemize}
\item
With the measurements $\left(\Gamma_{a_0\to \pi \eta} \times \Gamma_{a_0\to \gamma \gamma}\right)/\Gamma^{\rm tot} = 0.21 \, {\rm keV}$ and
$\Gamma_{a_0\to \gamma \gamma} = 0.30 \pm 0.10 \, {\rm keV}$ \cite{Amsler:1997up},
one get $\Gamma_{a_0 \to \pi \eta} = 0.053 \pm 0.018  \, {\rm GeV}$. We do not use eq. (\ref{eq:ga0KK-coup}) to determine the partial width since it is an approximation expression under the narrow width limit.
Furthermore, one can get $\Gamma_{a_0 \to K\bar K} = 0.009 \pm 0.003 \, {\rm GeV}$ with the measurement
$\Gamma_{a_0 \to K\bar K}/\Gamma_{a_0 \to \pi \eta}=0.177$ \cite{PDG-2020}.
\item
The partial decay widths of $a_0^\prime$ to $K\bar K$ and $\pi\eta$ states are decided by
the measured branching ratios $\Gamma_{{a_0^\prime}\to K\bar K}/\Gamma_{a_0^\prime}^{\rm tot} = 0.082\pm0.028$ and
$\Gamma_{{a_0^\prime}\to \pi \eta}/\Gamma_{a_0^\prime}^{\rm tot} = 0.093\pm0.020$ \cite{PDG-2020}.
\item
For the $a_0^{\prime\prime}$ decays, there is no direct measurement and the predictions from different models vary widely.
For example, the Extended Linear Sigma Model (eLSM) states that $\Gamma_{a_0^{\prime\prime}}\to K\bar K = 94 \pm 54 \, {\rm MeV}$ 
and $\Gamma_{a_0^{\prime\prime}} \to \pi\eta = 94 \pm 16 \, {\rm MeV}$ \cite{Parganlija:2016yxq},
while the $3^3P_0$ quark model gives the result $0.74 \, {\rm MeV}$ and $5.13 \, {\rm MeV}$ correspondently \cite{Wang:2017pxm}.
So in our evaluation, we take the largest interval of this variable to account its uncertainty.
\item
To close the descriptions, we summary the intervals of partial decay widths as
\beq
&&\Gamma_{a_0 \to K\bar K} = 0.009\pm 0.003 \, {\rm GeV}\,, \quad \quad \Gamma_{a_0 \to \pi \eta} = 0.053\pm 0.018 \, {\rm GeV} \,, \non
&&\Gamma_{a_0^\prime \to K\bar K} = 0.022 \pm 0.008 \, {\rm GeV} \,, \quad \quad  \Gamma_{a_0^\prime \to \pi \eta}=0.025 \pm 0.006 \, {\rm GeV} \,, \non
&&\Gamma_{a_0^{\prime\prime} \to K\bar K} \in [0, 0.150] \, {\rm GeV} \,, \hspace{1.7cm}
\Gamma_{{a_0^{\prime\prime}}\to \pi \eta} \in [0, 0.110] \, {\rm GeV} \,.
\label{eq:partial-width}
\eeq
\end{itemize}
Concerning the Gegenbauer expansion of scalar mesons,
we take into account the first two odd moments $B_1$ and $B_3$ in the twist 2 LCDAs \cite{Cheng:2005nb}
and the asymptotic terms in the twist 3 LCDAs due to the large theoretical uncertainty of $a_m$ and $b_m$
\cite{Lu:2006fr,Han:2013zg,Wang:2014vra}. They are
\beq
&&B_1^{a_0} = -0.93 \pm 0.10  \,, \quad B_3^{a_0} = 0.14 \pm 0.08 \,, \non
&&B_1^{a_0^\prime} = 0.89 \pm 0.20 \,, \quad\;\;\; B_3^{a_0^\prime} = -1.38 \pm 0.18 \,
\eeq
in the first scenario, and
\beq
&&B_1^{a_0^\prime} = -0.58 \pm 0.12 \,, \quad B_3^{a_0^\prime} = -0.49 \pm 0.15 \,, \non
&&B_1^{a_0^{\prime\prime}} = 0.73 \pm 0.45 \,, \quad\;\;\; B_3^{a_0^{\prime\prime}} = 0.17 \pm 0.20 \,
\eeq
in the second scenario, where the default scale at $1 \, {\rm GeV}$ is indicated.

\begin{table}[t]
\vspace{-2mm}
\caption{The PQCD predictions of branching fractions (in unit of $10^{-6}$) and ${\it CP}$ violations of
$B \to a_0 \left[ \to K\bar K/ \pi\eta \right] h$ decays in the first scenario of multiparticle configurations of $a_0$.}
\begin{center}
\begin{tabular}{l|c|r|r|c|r}
\toprule
{\rm Decay modes} & \; {\rm Quasi-two-body} \; & \; {\rm narrow approx.} \;
& \; {\rm two-body} \;\; & \; {\rm data} \cite{PDG-2020} \; &   {\rm CPV}~~~~~~~~~\non
\hline
$B^+ \to a_0^+ \left[\to K^+{\bar K}^0\right] \pi^0$ &$0.08^{+0.03+0.00}_{-0.03-0.00}$ &&$0.41^{+0.00}_{-0.23}$ \cite{Zhang:2010fcy} & &$38.2^{+3.5+3.5}_{-1.4-7.7}$ \non
$\quad\;\;\, \to a_0^+ \left[\to \pi^+ \eta\right] \pi^0$ &$0.37^{+0.14+0.04}_{-0.08-0.04}$&$0.52^{+0.20+0.06}_{-0.11-0.05}$ \;\;  & $0.70^{+0.32}_{-0.23}$ \cite{Cheng:2013fba} & $< 1.4$ &$56.3^{+1.2+2.8}_{-3.1-7.3}$ \non     
$B^+ \to a_0^0 \left[\to K^-K^+\right] \pi^+$ &$0.33^{+0.12+0.04}_{-0.08-0.04}$ &&$2.8^{+0.0}_{-1.3}$ \cite{Zhang:2010fcy} & &$24.1^{+2.6+6.5}_{-2.4-6.6}$ \non
$\quad\;\;\, \to a_0^0 \left[\to \pi^0 \eta\right] \pi^+$ &$2.41^{+0.91+0.37}_{-0.62-0.30}$ &$3.44^{+1.29+0.54}_{-0.88-0.42}$ \;\;  & $4.9^{+1.4}_{-1.3}$ \cite{Cheng:2013fba} & $< 5.8$ &$26.5^{+0.1+5.4}_{-2.7-6.1}$ \non
$B^+ \to a_0^+ \left[\to K^+{\bar K}^0\right] K^0$ &$0.26^{+0.03+0.16}_{-0.01-0.10}$ &&$6.9^{+2.4}_{-2.1}$  \cite{Shen:2006ms} & &$6.1^{+5.5+5.4}_{-4.9-6.2}$ \non
$\quad\;\;\, \to a_0^+ \left[\to \pi^+ \eta\right] K^0$ &$0.94^{+0.04+0.85}_{-0.02-0.51}$ &$1.35^{+0.06+1.21}_{-0.03-0.72}$  \;\; & $0.08^{+2.20}_{-0.11}$ \cite{Cheng:2013fba} & $<3.9$ &$3.72^{+2.4+5.1}_{-3.3-3.0}$ \non
$B^+ \to a_0^0 \left[\to K^-K^+\right] K^+$ &$0.11^{+0.0+0.06}_{-0.0-0.04}$ &&$3.5^{+1.1}_{-1.2}$ \cite{Shen:2006ms} & &$-26.4^{+4.8+4.9}_{-4.2-6.7}$ \non
$\quad\;\;\, \to a_0^0 \left[\to \pi^0 \eta\right] K^+$ &$1.06^{+0.02+0.59}_{-0.04-0.42}$ &$1.51^{+0.03+0.85}_{-0.06-0.61}$ \;\;  & $0.34^{+1.12}_{-0.16}$ \cite{Cheng:2013fba} & $<2.5$ &$-21.3^{+4.0+7.4}_{-4.6-9.5}$ \non
\hline
$B^0 \to a_0^+ \left[\to K^+{\bar K}^0\right] \pi^-$ &$0.17^{+0.06+0.01}_{-0.04-0.01}$ &&$0.51^{+0.12}_{-0.12}$  \cite{Zhang:2010fcy} & &$70.5^{+0.5+6.9}_{-3.1-7.4}$ \non
$\quad\;\, \to a_0^+ \left[\to \pi^+ \eta\right] \pi^-$ &$0.67^{+0.24+0.06}_{-0.15-0.07}$ &$0.95^{+0.34+0.08}_{-0.22-0.10}$  \;\; &$0.58^{+0.65}_{-0.25}$ \cite{Cheng:2013fba} & &$68.3^{+3.4+6.4}_{-6.2-7.2}$\non
$B^0 \to a_0^0 \left[\to K^-K^+\right] \pi^0$ &$0.04^{+0.02+0.01}_{-0.01-0.00}$ &&$0.51^{+0.12}_{-0.11}$ \cite{Zhang:2010fcy} & &$79.4^{+0.4+7.9}_{-6.6-9.6}$ \non
$\quad\;\, \to a_0^0 \left[\to \pi^0 \eta\right] \pi^0$ &$0.33^{+0.09+0.05}_{-0.05-0.06}$ &$0.47^{+0.12+0.07}_{-0.07-0.08}$  \;\; &$1.0^{+0.5}_{-0.3}$ \cite{Cheng:2013fba} & &$84.1^{+7.2+1.9}_{-5.6-5.9}$ \non
$B^0 \to a_0^- \left[\to K^-K^0\right] \pi^+$ &$3.48^{+1.33+0.34}_{-0.92-0.29}$ &&$0.86^{+0.17}_{-0.17}$ \cite{Zhang:2010fcy} & &$17.8^{+2.3+3.1}_{-2.2-3.5}$ \non
$\quad\;\, \to a_0^- \left[\to \pi^- \eta\right] \pi^+$ &$14.8^{+5.6+1.7}_{-3.9-1.4}$ &$21.1^{+7.9+2.3}_{-5.6-2.1}$ \;\;  &$5.3^{+1.7}_{-1.4}$ \cite{Cheng:2013fba} & &$20.6^{+2.6+2.7}_{-2.7-3.6}$\non
$B^0 \to a_0^0 \left[\to K^-K^+\right] K^0$ &$0.11^{+0.03+0.04}_{-0.01-0.02}$ &&$4.7^{+1.4}_{-1.4}$ \cite{Shen:2006ms} & &$-27.5^{+6.9+5.6}_{-1.7-2.1}$ \non
$\quad\;\, \to a_0^0 \left[\to \pi^0 \eta\right] K^0$ &$1.36^{+0.21+0.43}_{-0.23-0.51}$ &$1.95^{+0.30+0.61}_{-0.32-0.72}$  \;\; & $0.05^{+0.91}_{-0.05}$ \cite{Cheng:2013fba} & $<7.8$ &$-43.2^{+1.7+5.7}_{-7.8-8.8}$ \non
$B^0 \to a_0^- \left[\to K^-K^0\right] K^+$ &$0.99^{+0.14+0.38}_{-0.09-0.33}$ &&$9.7^{+3.3}_{-2.8}$ \cite{Shen:2006ms} & &$-69.7^{+1.2+1.7}_{-4.1-2.6}$ \non
$\quad\;\, \to a_0^- \left[\to \pi^- \eta\right] K^+$ &$4.51^{+0.60+1.72}_{-0.61-1.60}$ &$6.44^{+0.85+2.53}_{-0.87-2.33}$ \;\;  & $0.34^{+2.35}_{-0.14}$ \cite{Cheng:2013fba} & $<1.9$ &$-83.2^{+2.5+3.4}_{-9.6-9.9}$ \non
\toprule
\end{tabular}
\end{center}
\label{table2}
\end{table}

Our numerical evaluations are carried out in two scenarios.
In the first scenario, we treat $a_0$ as the lowest-lying $q{\bar q}$ state and $a_0^\prime$ as its first excited state,
and study the contributions from $a_0$ and $a_0^\prime$ in the $B \to a_0^{(\prime)} \left[ \to K\bar K/\pi\eta \right] h$ decays.
The second scenario indicates that $a_0^\prime$ is the lowest-lying $q{\bar q}$ state and $a_0^{\prime\prime}$ is the first excited state,
with this ansatz we study their contributions in the $B \to a_0^{\prime/\prime\prime} \left[ \to K\bar K/\pi\eta \right] h$ decays.

\begin{table}[t]
\vspace{-2mm}
\caption{The same as table \ref{table2}, but for the $B \to a^\prime_0 \left[ \to K\bar K/ \pi\eta \right] h$ decays.}
\begin{center}
\begin{tabular}{l|c|c|c|r}
\toprule
{\rm Decay modes} & \; {\rm Quasi-two-body} \; & \; {\rm narrow approx.} \; & \; {\rm two-body} \cite{Cheng:2013fba} \; & \quad {\rm CPV}~~~~~~~~~ \non
\hline
$B^+ \to a_0^{\prime +} \left[\to K^+{\bar K}^0\right] \pi^0$ &$0.08^{+0.01+0.00}_{-0.02-0.01}$ &$0.94^{+0.12+0.03}_{-0.20-0.19}$ & &$-4.6^{+1.9+4.5}_{-9.2-9.1}$ \non
$\quad\;\;\, \to a_0^{\prime +} \left[\to \pi^+ \eta\right] \pi^0$ &$0.09^{+0.02+0.01}_{-0.02-0.01}$ &$0.95^{+0.18+0.06}_{-0.21-0.16}$ & $0.4^{+0.3}_{-0.3}$ &$-13.2^{+9.1+7.8}_{-12.7-8.3}$\non
$B^+ \to a_0^{\prime 0} \left[\to K^-K^+\right] \pi^+$ &$0.12^{+0.04+0.03}_{-0.03-0.02}$ &$2.81^{+1.09+0.73}_{-0.61-0.54}$ & &$26.6^{+23.7+13.3}_{-19.9-17.0}$ \non
$\quad\;\;\, \to a_0^{\prime 0} \left[\to \pi^0 \eta\right] \pi^+$ &$0.28^{+0.10+0.07}_{-0.07-0.04}$ &$3.02^{+1.09+0.78}_{-0.74-0.52}$ & $2.7^{+0.7}_{-0.7}$ &$28.2^{+14.6+13.0}_{-16.7-18.4}$ \non
$B^+ \to a_0^{\prime +} \left[\to K^+{\bar K}^0\right] K^0$ &$1.28^{+0.03+0.45}_{-0.05-0.40}$ &$15.6^{+0.4+2.7}_{-0.6-2.2}$ & &$4.8^{+1.5+2.9}_{-0.5-3.1}$ \non
$\quad\;\;\, \to a_0^{\prime +} \left[\to \pi^+ \eta\right] K^0$ &$1.50^{+0.04+0.53}_{-0.06-0.48}$ &$16.1^{+0.5+5.7}_{-0.5-5.0}$ & $2.7^{+10.1}_{-3.2}$ &$4.8^{+1.4+1.9}_{-0.3-0.6}$\non
$B^+ \to a_0^{\prime 0} \left[\to K^-K^+\right] K^+$ &$0.44^{+0.01+0.16}_{-0.01-0.14}$ &$10.8^{+0.3+2.9}_{-0.4-2.6}$ & &$1.0^{+0.1+5.9}_{-0.8-3.2}$ \non
$\quad\;\;\, \to a_0^{\prime 0} \left[\to \pi^0 \eta\right] K^+$ &$1.02^{+0.03+0.37}_{-0.03-0.34}$ &$11.1^{+0.2+4.0}_{-0.4-3.6}$ & $0.7^{+3.2}_{-0.6}$ &$0.8^{+0.5+6.2}_{-0.6-3.0}$\non
\hline
$B^0 \to a_0^{\prime +} \left[\to K^+{\bar K}^0\right] \pi^-$ &$0.04^{+0.01+0.01}_{-0.01-0.01}$ &$0.49^{+0.14+0.17}_{-0.08-0.14}$ & &$-24.0^{+12.0+19.3}_{-13.0-18.2}$ \non
$\quad\;\, \to a_0^{\prime +} \left[\to \pi^+ \eta\right] \pi^-$ &$0.03^{+0.01+0.01}_{-0.00-0.01}$ &$0.36^{+0.10+0.17}_{-0.03-0.12}$ & $0.02^{+0.75}_{-0.01}$ &$-20.7^{+15.6+25.4}_{-10.7-23.0}$\non
$B^0 \to a_0^{\prime 0} \left[\to K^-K^+\right] \pi^0$ &$0.03^{+0.01+0.01}_{-0.01-0.01}$ &$0.67^{+0.16+0.21}_{-0.12-0.15}$ & &$-22.0^{+19.1+18.0}_{-13.1-19.6}$ \non
$\quad\;\, \to a_0^{\prime 0} \left[\to \pi^0 \eta\right] \pi^0$ &$0.07^{+0.01+0.02}_{-0.01-0.02}$ &$0.70^{+0.16+0.18}_{-0.08-0.23}$ & $1.3^{+2.1}_{-1.1}$ &$-31.9^{+13.4+19.1}_{-8.5-19.2}$ \non
$B^0 \to a_0^{\prime -} \left[\to K^-K^0\right] \pi^+$ &$1.08^{+0.35+0.18}_{-0.24-0.17}$ &$13.2^{+4.2+2.2}_{-2.9-2.2}$ & &$24.8^{+1.2+4.4}_{-0.7-5.3}$ \non
$\quad\;\, \to a_0^{\prime -} \left[\to \pi^- \eta\right] \pi^+$ &$1.24^{+0.39+0.21}_{-0.27-0.21}$ &$13.3^{+4.3+2.4}_{-2.9-2.2}$ & $11.2^{+5.2}_{-5.7}$ &$25.6^{+2.9+4.3}_{-0.8-5.9}$\non
$B^0 \to a_0^{\prime 0} \left[\to K^-K^+\right] K^0$ &$0.25^{+0.02+0.12}_{-0.02-0.09}$ &$6.06^{+0.50+3.01}_{-0.56-2.28}$ & &$-0.3^{+2.9+2.5}_{-4.9-0.4}$ \non
$\quad\;\, \to a_0^{\prime 0} \left[\to \pi^0 \eta\right] K^0$ &$0.58^{+0.05+0.29}_{-0.05-0.22}$ &$6.27^{+0.52+3.14}_{-0.52-2.39}$ & $0.9^{+3.8}_{-1.1}$ &$-0.6^{+3.4+2.1}_{-6.5-0.8}$\non
$B^0 \to a_0^{\prime -} \left[\to K^-K^0\right] K^+$ &$2.62^{+0.31+0.71}_{-0.29-0.62}$ &$32.0^{+3.8+6.7}_{-3.4-4.7}$ & &$-18.9^{+2.7+1.1}_{-2.3-3.8}$ \non
$\quad\;\, \to a_0^{\prime -} \left[\to \pi^- \eta\right] K^+$ &$3.04^{+0.35+0.80}_{-0.33-0.73}$ &$32.7^{+3.7+7.7}_{-3.6-6.6}$ & $1.9^{+8.1}_{-1.8}$ &$-19.5^{+2.0+0.9}_{-2.2-4.0}$\non
\toprule
\end{tabular}
\end{center}
\label{table3}
\end{table}

In table \ref{table2} and table \ref{table3}, we present the PQCD predictions of of $B \to a_0 \left[ \to K\bar K/ \pi\eta \right] h$
and $B \to a_0^\prime \left[ \to K\bar K/ \pi\eta \right] h$ decays in the first scenario of multiparticle configurations of $a_0$, respectively.
Besides the result of quasi-two-body decays, saying the branching fractions (in the 2nd column) and the ${\it CP}$ violations (in the last column),
we list the branching fractions of two-body $B \to a_0^{\prime} h$ decays\footnote{The narrow width approximation is not applicable 
to the modes involving $a_0 h \to KK$ due to the threshold effect, so in table \ref{table2} we do not list the result of two-body $B \to a_0 h$ decay}
obtained in the narrow width approximation (in the 3rd column),
for the sake of comparison, the direct two-body calculations based on PQCD \cite{Shen:2006ms} and QCDF approach \cite{Cheng:2013fba},
and also the available data are list too (in the 4th and 5th columns).
The theoretical uncertainties come from the inputs of LCDAs, 
mainly from the inverse moment $\omega_B$ which we put as the first error source, 
the uncertainties from Gegenbauer moments $B^{a_0}_1, B^{a_0}_3$ of dimeson systems are add together as the second error, 
we do not consider the uncertainty from other parameters, like $f_{a_0}, {\bar f}_{a_0}$ since their influences are small. 
We comment in orders,
\begin{itemize}
\item[(a)] The branching fractions of quasi-two-body channels with strong decays $a_0 \to \pi\eta$
is about 5 times larger than that with the strong decay $a_0 \to K\bar K$,
which is understood by the suppressed phase space for $K\bar K$ state.
\vspace{-4mm}
\item[(b)] Under the narrow width approximation of the quasi-two-body decays,
we extract the branching fractions of relevant two body decays $B \to a_0^{(\prime)} h$.
The result obtained from the $a_0^\prime \to K\bar K$ and $a_0^\prime \to \pi\eta$ modes are consist with each other with in the uncertainties,
more important is that this result have a large discrepancy with the direct two-body calculation from PQCD \cite{Shen:2006ms}
and QCDF \cite{Cheng:2013fba}, revealing the important role of width effects of $a_0$ and $a_0^\prime$.
\vspace{-4mm}
\item[(c)] In the $B \to a_0^{\prime} h$ and the following $B \to a_0^{\prime\prime} h$ decays, 
only the partial width expression is used due to the lacking of direct measurements,
that's why the branching fractions of these decays extracted from $K\bar K$ and $\pi\eta$ modes are very close to each other.
\vspace{-4mm}
\item[(d)] The PQCD predictions of branching fractions of the six $B \to a_0^{(+,0)} \left[ \to \pi\eta \right] h$ quasi-two-body decays
do not excess the experimental upper limit, 
the predictions of two channels $B^0 \to a_0^{\pm} \left[ \to \pi^\pm \eta\right] \pi^\mp$ 
excess the experimental upper limit $3.1 \times 10^{-6}$ \cite{PDG-2020} at the first glance, 
but the large uncertainties would be more larger if we considering the uncertainty of $\omega_B = 440 \pm 110 \, {\rm MeV}$. 
So with in acceptable limits, the $q{\bar q}$ configuration of $a_0$ is still survival in $B$ decays. 
It is shown that the decaying channel $B^0 \to a_0^- \left[ \to \pi^- \eta \right] \pi^+$ has the largest branching fraction,
and we suggest the measurement to examine the $q{\bar q}$ configuration.
\end{itemize}
\begin{table}[t]
\vspace{-2mm}
\caption{The PQCD predictions of branching fractions (in unit of $10^{-6}$) and ${\it CP}$ violations of
$B \to a^\prime_0 \left[ \to K\bar K/ \pi\eta \right] h$ decays in the second scenario of multiparticle configurations of $a_0$.}
\begin{center}
\begin{tabular}{l|c|r|c|r}
\toprule
{\rm Decay modes} \; & \; {\rm Quasi-two-body} \; & \; {\rm narrow approx.} \; & \; {\rm two-body} \cite{Cheng:2013fba} \; & \quad  {\rm CPV}~~~~~~~~~ \non
\hline
$B^+ \to a_0^{\prime +} \left[\to K^+{\bar K}^0\right] \pi^0$ &$0.10^{+0.04+0.00}_{-0.03-0.01}$ &$1.24^{+0.52+0.03}_{-0.34-0.09}$ \;\; & &$-19.2^{+4.9+5.1}_{-4.3-8.1}$ \non
$\quad\;\;\, \to a_0^{\prime +} \left[\to \pi^+ \eta\right] \pi^0$ &$0.12^{+0.05+0.01}_{-0.03-0.01}$ &$1.24^{+0.53+0.08}_{-0.31-0.10}$ \;\; & $2.1^{+1.1}_{-0.8}$ &$-15.2^{+2.4+5.3}_{-3.1-6.8}$\non
$B^+ \to a_0^{\prime 0} \left[\to K^-K^+\right] \pi^+$ &$0.25^{+0.11+0.04}_{-0.07-0.04}$ &$6.07^{+2.80+1.06}_{-1.77-0.98}$ \;\; & &$-0.1^{+1.6+3.6}_{-1.5-2.0}$ \non
$\quad\;\;\, \to a_0^{\prime 0} \left[\to \pi^0 \eta\right] \pi^+$ &$0.56^{+0.27+0.09}_{-0.16-0.09}$ &$6.01^{+2.91+1.08}_{-1.72-0.99}$ \;\; & $5.1^{+1.8}_{-1.7}$ &$1.0^{+0.8+3.4}_{-2.6-3.7}$ \non
$B^+ \to a_0^{\prime +} \left[\to K^+{\bar K}^0\right] K^0$ &$1.29^{+0.03+0.68}_{-0.02-0.52}$ &$15.8^{+0.4+4.6}_{-0.2-3.9}$ \;\; & &$0.5^{+0.1+0.6}_{-0.1-0.6}$ \non
$\quad\;\;\, \to a_0^{\prime +} \left[\to \pi^+ \eta\right] K^0$ &$1.51^{+0.04+0.79}_{-0.03-0.62}$ &$16.3^{+0.5+8.4}_{-0.3-6.5}$ \;\; & $4.2^{+18.8}_{-4.8}$ &$0.3^{+0.2+0.8}_{-0.3-0.9}$\non
$B^+ \to a_0^{\prime 0} \left[\to K^-K^+\right] K^+$ &$0.50^{+0.00+0.23}_{-0.01-0.20}$ &$12.3^{+0.1+3.9}_{-0.3-3.1}$ \;\; & &$-22.7^{+2.4+1.3}_{-3.2-0.8}$ \non
$\quad\;\;\, \to a_0^{\prime 0} \left[\to \pi^0 \eta\right] K^+$ &$1.13^{+0.00+0.56}_{-0.02-0.43}$ &$12.2^{+0.1+5.9}_{-0.1-4.7}$ \;\; & $2.2^{+8.1}_{-2.2}$ &$-23.6^{+3.0+3.8}_{-2.0-1.8}$\non
\hline
$B^0 \to a_0^{\prime +} \left[\to K^+{\bar K}^0\right] \pi^-$ &$0.13^{+0.05+0.01}_{-0.03-0.01}$ &$1.56^{+0.57+0.18}_{-0.37-0.13}$ \;\; & &$24.8^{+0.5+6.5}_{-0.6-6.8}$ \non
$\quad\;\, \to a_0^{\prime +} \left[\to \pi^+ \eta\right] \pi^-$ &$0.14^{+0.05+0.01}_{-0.03-0.01}$ &$1.51^{+0.57+0.16}_{-0.37-0.12}$ \;\; & $0.74^{+2.9}_{-0.6}$ &$28.5^{+0.4+6.7}_{-0.5-4.6}$\non
$B^0 \to a_0^{\prime 0} \left[\to K^-K^+\right] \pi^0$ &$0.05^{+0.01+0.01}_{-0.01-0.01}$ &$1.07^{+0.16+0.34}_{-0.10-0.32}$ \;\; & &$26.1^{+5.1+8.1}_{-6.7-8.6}$ \non
$\quad\;\, \to a_0^{\prime 0} \left[\to \pi^0 \eta\right] \pi^0$ &$0.10^{+0.01+0.04}_{-0.01-0.03}$ &$1.10^{+0.15+0.34}_{-0.11-0.34}$ \;\; & $3.3^{+3.1}_{-1.7}$ &$24.3^{+6.5+17.2}_{-6.4-12.3}$\non
$B^0 \to a_0^{\prime -} \left[\to K^-K^0\right] \pi^+$ &$3.61^{+1.32+0.38}_{-0.92-0.36}$ &$44.0^{+16.2+4.7}_{-11.2-4.4}$ \;\; & &$25.8^{+3.3+4.0}_{-3.0-3.6}$ \non
$\quad\;\, \to a_0^{\prime -} \left[\to \pi^- \eta \right] \pi^+$ &$4.15^{+1.52+0.45}_{-1.05-0.42}$ &$44.6^{+16.4+4.9}_{-11.4-4.5}$ \;\; & $2.5^{+3.8}_{-1.0}$ &$26.1^{+3.3+3.5}_{-2.9-3.6}$ \non
$B^0 \to a_0^{\prime 0} \left[\to K^-K^+\right] K^0$ &$0.33^{+0.01+0.17}_{-0.00-0.13}$ &$8.10^{+0.16+4.05}_{-0.02-3.01}$ \;\; & &$-6.3^{+0.1+0.8}_{-2.5-3.1}$ \non
$\quad\;\, \to a_0^{\prime 0} \left[\to \pi^0 \eta\right] K^0$ &$0.78^{+0.01+0.40}_{-0.00-0.29}$ &$8.34^{+0.17+4.29}_{-0.02-3.96}$ \;\; & $1.9^{+7.8}_{-2.2}$ &$-7.5^{+0.7+1.1}_{-2.3-2.4}$ \non
$B^0 \to a_0^{\prime -} \left[\to K^-K^0\right] K^+$ &$2.93^{+0.49+1.05}_{-0.35-0.89}$ &$35.7^{+6.0+9.9}_{-4.2-9.7}$ \;\; & &$-46.7^{+1.6+4.1}_{-0.3-3.6}$ \non
$\quad\;\, \to a_0^{\prime -} \left[\to \pi^- \eta\right] K^+$ &$3.39^{+0.52+1.02}_{-0.39-1.02}$ &$36.5^{+5.6+13.0}_{-4.2-10.0}$ \;\; & $3.5^{+17.5}_{-3.9}$ &$-46.0^{+3.3+4.0}_{-1.5-4.1}$ \non
\toprule
\end{tabular}
\end{center}
\label{table4}
\end{table}

\begin{table}[t]
\vspace{-2mm}
\caption{The same as table \ref{table4}, but for the $B \to a^{\prime\prime}_0 \left[ \to K\bar K/ \pi\eta\right] h$ decays.}
\begin{center}
\begin{tabular}{l|c|r|r}
\toprule
{\rm Decay modes} \; & \; {\rm Quasi-two-body} \; & \; {\rm narrow approx.} \;   & \quad {\rm CPV}~~~~~~~~~ \non
\hline
$B^+ \to a_0^{\prime\prime +}  \left[\to K^+{\bar K}^0\right] \pi^0$ &$0.38^{+0.17+0.02}_{-0.10-0.02}\pm 0.22$   &$1.14^{+0.50+0.06}_{-0.30-0.04}$ \;\;  &$17.6^{+1.9+0.8}_{-4.1-0.8}$ \non
$\quad\;\;\, \to a_0^{\prime\prime +}  \left[\to \pi^+ \eta\right] \pi^0$ &$0.39^{+0.16+0.01}_{-0.10-0.01}\pm 0.07$  &$1.16^{+0.48+0.03}_{-0.31-0.04}$ \;\;  &$13.3^{+3.1+0.9}_{-0.1-0.7}$\non
$B^+ \to a_0^{\prime\prime 0}  \left[\to K^-K^+\right] \pi^+$ &$3.04^{+1.22+0.59}_{-0.81-0.48}\pm 1.75$  &$18.1^{+7.3+3.5}_{-4.8-2.8}$ \;\;  &$-6.7^{+1.1+2.1}_{-1.3-2.2}$ \non
$\quad\;\;\, \to a_0^{\prime\prime 0}  \left[\to \pi^0 \eta\right] \pi^+$ &$6.31^{+2.48+1.20}_{-0.71-1.07}\pm 1.07$  &$18.8^{+7.4+3.5}_{-5.1-3.3}$ \;\;  &$-7.0^{+0.9+0.9}_{-0.9-1.7}$ \non
$B^+ \to a_0^{\prime\prime +}  \left[\to K^+{\bar K}^0\right] K^0$ &$2.60^{+0.15+3.18}_{-0.08-2.01}\pm 1.49$  &$7.73^{+0.37+9.49}_{-0.25-6.01}$ \;\; &$0.6^{+0.0+1.4}_{-0.7-0.6}$ \non
$\quad\;\;\, \to a_0^{\prime\prime +}  \left[\to \pi^+ \eta\right] K^0$ &$2.62^{+0.15+3.21}_{-0.09-2.04}\pm 0.44$  &$7.81^{+0.44+9.59}_{-0.25-6.09}$ \;\; &$0.5^{+0.4+0.7}_{-0.8-0.3}$ \non
$B^+ \to a_0^{\prime\prime 0}  \left[\to K^-K^+\right] K^+$ &$0.58^{+0.00+0.83}_{-0.00-0.52}\pm 0.33$  &$3.46^{+0.02+4.96}_{-0.01-3.13}$ \;\; &$-27.9^{+3.8+6.3}_{-4.2-6.8}$ \non
$\quad\;\;\, \to a_0^{\prime\prime 0}  \left[\to \pi^0 \eta\right] K^+$ &$1.19^{+0.01+1.66}_{-0.00-1.07}\pm 0.20$  &$3.56^{+0.01+4.95}_{-0.05-3.13}$ \;\;  &$-30.8^{+4.4+7.3}_{-5.4-7.0}$ \non
\hline
$B^0 \to a_0^{\prime\prime +}  \left[\to K^+{\bar K}^0\right] \pi^-$ &$1.02^{+0.35+0.27}_{-0.23-0.21}\pm 0.59$  &$3.05^{+1.03+0.80}_{-0.72-0.66}$ \;\; &$-7.0^{+2.5+8.4}_{-2.6-8.5}$ \non
$\quad\;\, \to a_0^{\prime\prime +}  \left[\to \pi^+ \eta\right] \pi^-$ &$1.01^{+0.35+0.27}_{-0.23-0.20}\pm 0.17$  &$3.02^{+1.04+0.83}_{-0.70-0.64}$ \;\;  &$-7.8^{+3.2+8.8}_{-2.2-7.9}$ \non
$B^0 \to a_0^{\prime\prime 0}  \left[\to K^-K^+\right] \pi^0$ &$0.22^{+0.06+0.11}_{-0.05-0.10}\pm 0.13$  &$1.32^{+0.33+0.69}_{-0.27-0.59}$ \;\; &$-31.3^{+1.2+5.7}_{-2.9-8.9}$ \non
$\quad\;\, \to a_0^{\prime\prime 0}  \left[\to \pi^0 \eta\right] \pi^0$ &$0.44^{+0.12+0.24}_{-0.08-0.19}\pm 0.07$  &$1.30^{+0.36+0.72}_{-0.25-0.56}$ \;\; &$-32.6^{+2.7+7.5}_{-1.8-8.1}$ \non
$B^0 \to a_0^{\prime\prime -}  \left[\to K^-K^0\right] \pi^+$ &$4.76^{+2.03+1.45}_{-1.34-1.19}\pm 2.73$  &$14.2^{+8.0+4.2}_{-4.0-3.7}$ \;\; &$-24.2^{+3.6+11.5}_{-3.9-10.7}$ \non
$\quad\;\, \to a_0^{\prime\prime -}  \left[\to \pi^- \eta\right] \pi^+$ &$4.76^{+2.04+1.46}_{-1.34-1.19}\pm 0.81$  &$14.2^{+8.0+4.2}_{-4.1-3.7}$ \;\; &$-24.3^{+3.5+9.7}_{-4.0-10.1}$ \non
$B^0 \to a_0^{\prime\prime 0}  \left[\to K^-K^+\right] K^0$ &$0.86^{+0.22+0.73}_{-0.13-0.67}\pm 0.50$  &$5.15^{+1.33+4.40}_{-0.80-3.05}$ \;\; &$-1.8^{+1.4+0.9}_{-0.6-1.8}$ \non
$\quad\;\, \to a_0^{\prime\prime 0}  \left[\to \pi^0 \eta\right] K^0$ &$1.74^{+0.43+1.48}_{-0.26-1.04}\pm 0.29$  &$5.20^{+1.29+4.42}_{-0.79-3.12}$ \;\; &$-2.1^{+1.2+0.2}_{-0.2-1.1}$ \non
$B^0 \to a_0^{\prime\prime -}  \left[\to K^-K^0\right] K^+$ &$3.82^{+1.01+2.11}_{-0.67-1.32}\pm 2.19$  &$11.4^{+2.9+6.2}_{-2.0-3.9}$ \;\; &$24.9^{+0.7+5.5}_{-0.4-4.0}$ \non
$\quad\;\, \to a_0^{\prime\prime -}  \left[\to \pi^- \eta\right] K^+$ &$3.80^{+1.02+2.18}_{-0.64-1.31}\pm 0.65$  &$11.3^{+3.1+6.7}_{-1.8-3.8}$ \;\; &$25.6^{+1.2+8.5}_{-0.6-9.7}$ \non
\toprule
\end{tabular}
\end{center}
\label{table5}
\end{table}

We list in table \ref{table4} and table \ref{table5} with the PQCD predictions of
$B \to a^{\prime}_0 \left[ \to K\bar K/ \pi\eta \right] h$ and $B \to a^{\prime\prime}_0 \left[ \to K\bar K/ \pi\eta \right] h$ decays
in the second scenario of multiparticle configurations of $a_0$, respectively. 
For the later one, we also present the uncertainty (as the third error) in the quasi-two-body decays from the partial decay width 
$\Gamma_{a_0^{\prime\prime} \to K\bar K/\pi \eta}$ as demonstrated in Eq. (\ref{eq:partial-width}), 
this parameter would not bring additional uncertainty to the two-body decays under narrow approximation. 
Similar result is obtained with showing that the decaying channels $B^0 \to a_0^{\prime -}\left[ \to K^-K^0/\pi^-\eta \right] h$
have the largest branching fractions both for the quasi-two-body and the extracted two-body decays.
We would like to mark that our predictions of the $a^\prime_0$ contributions are comparable in the most of $B \to K\bar K h, \pi\eta h$ decays
no matter what's the scenarios of $a_0$ configurations are taken,
while for the channels $B^0 \to a_0^{\prime +} \left[ \to K^+{\bar K}^0, \pi^+\eta \right] \pi^-$
and $B^0 \to a_0^{\prime -} \left[ \to K^-K^0, \pi^-\eta \right] \pi^+$,
the predictions of branching fractions in the second scenario
are about three time larger in magnitude than that predicted in the first scenario\footnote{The PQCD predictions
in the second scenario for these channels consist with the result from factorisation approach under $SU(3)$ symmetry \cite{Li:2014oca},
and the predictions in both two scenarios are under the experiment upper limit.},
which provide another opportunity to check which one is the right with the future measurement.
In these tables we also list ${\it CP}$ violations which provide another observables to study
the interactions between different operators and/or topological amplitudes, especially the different sources of strong phases.

\begin{figure}[t]
\vspace{-1cm}
\centerline{\includegraphics[width=0.55\textwidth]{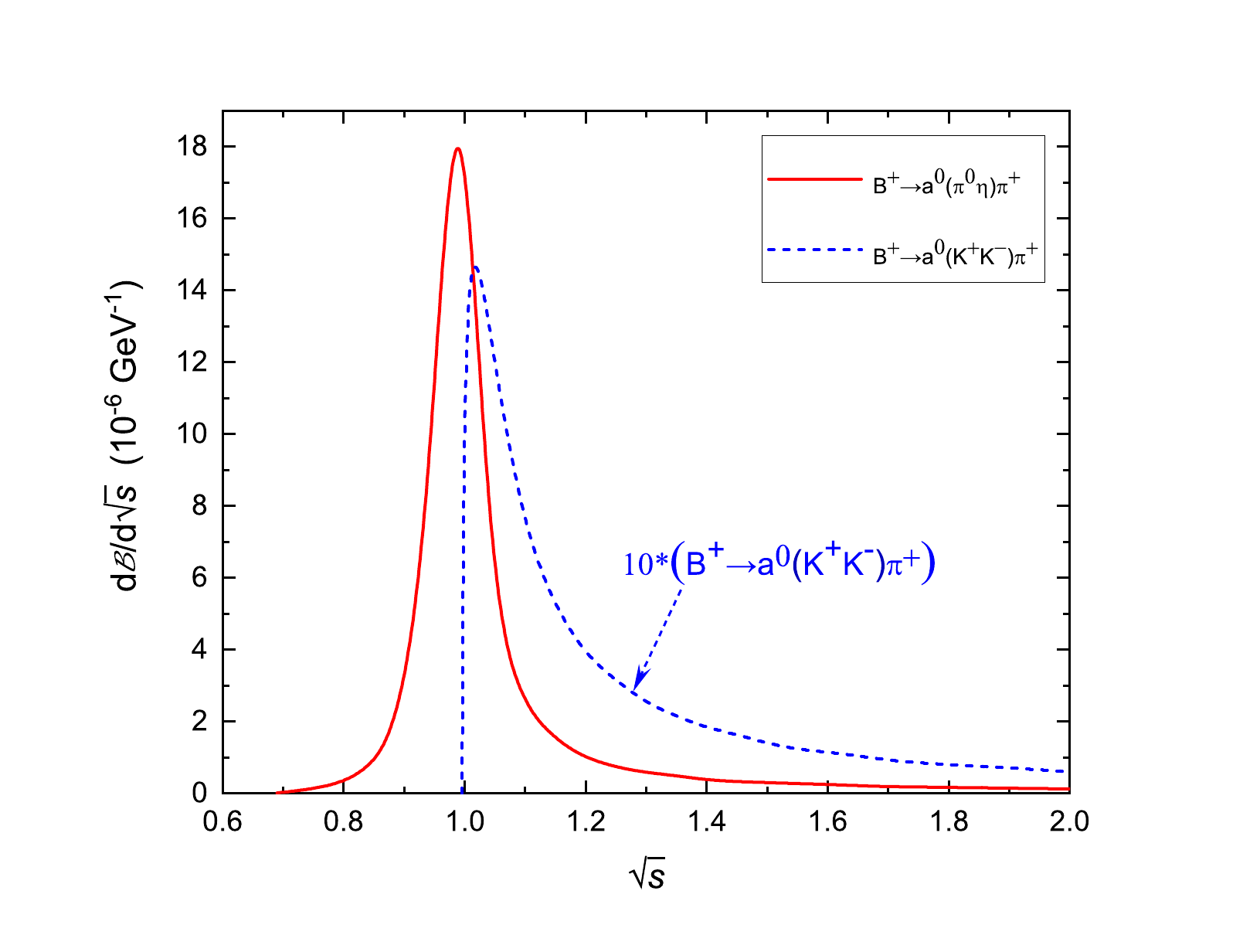}
\hspace{-1.0cm}
\includegraphics[width=0.55\textwidth]{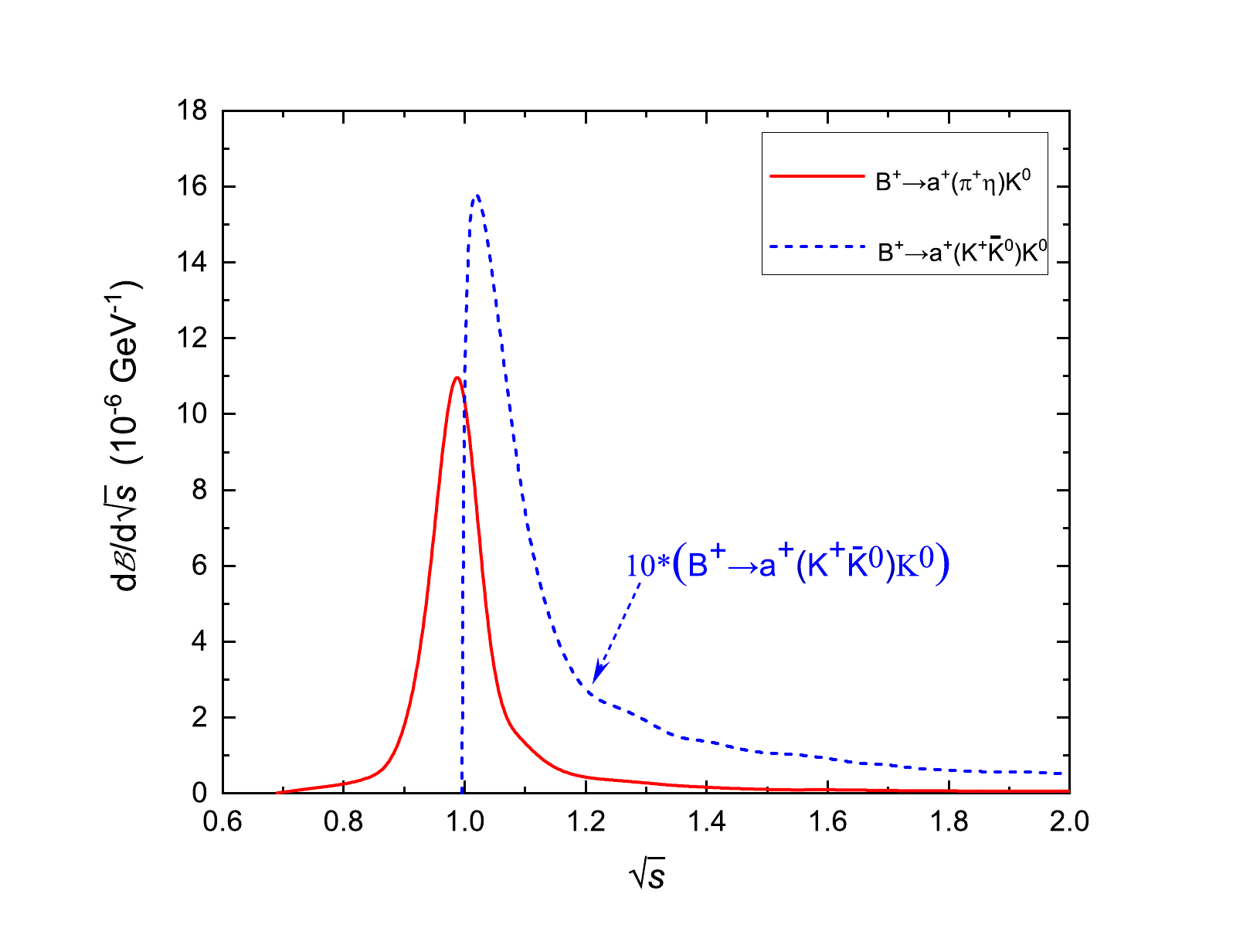} }
\vspace{-6mm}
\centerline{\includegraphics[width=0.55\textwidth]{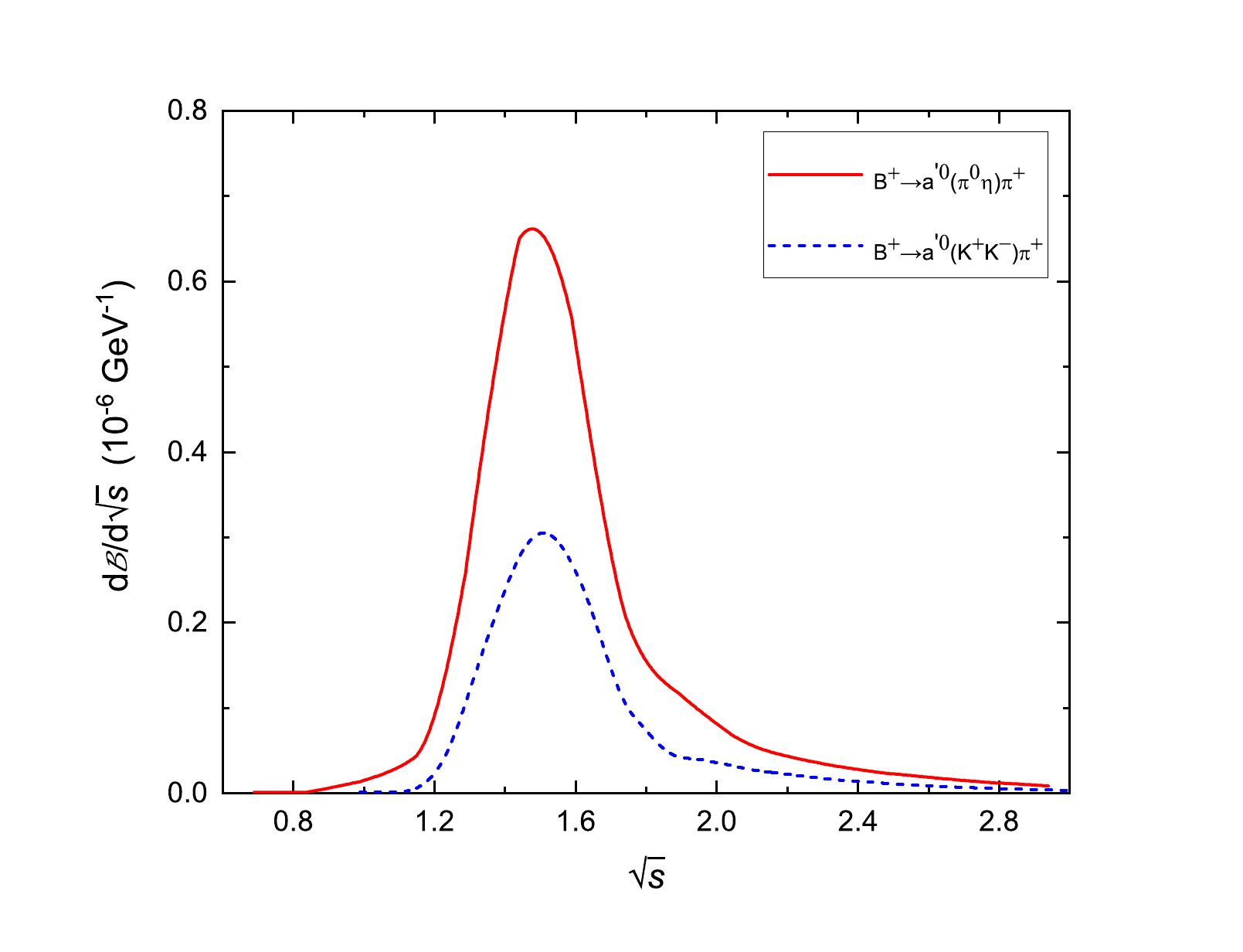}
\hspace{-1.0cm}
\includegraphics[width=0.55\textwidth]{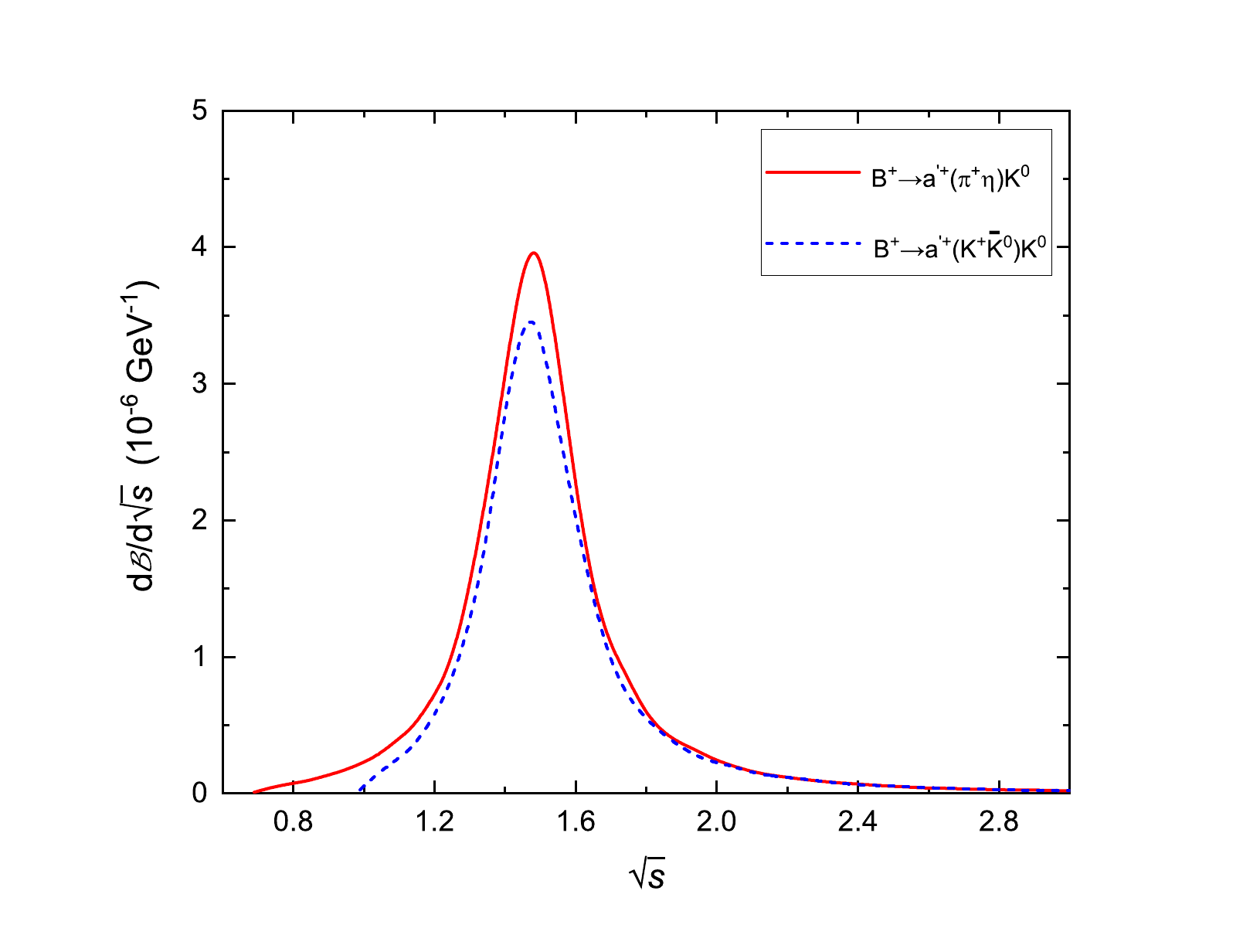} }
\vspace{-6mm}
\centerline{\includegraphics[width=0.55\textwidth]{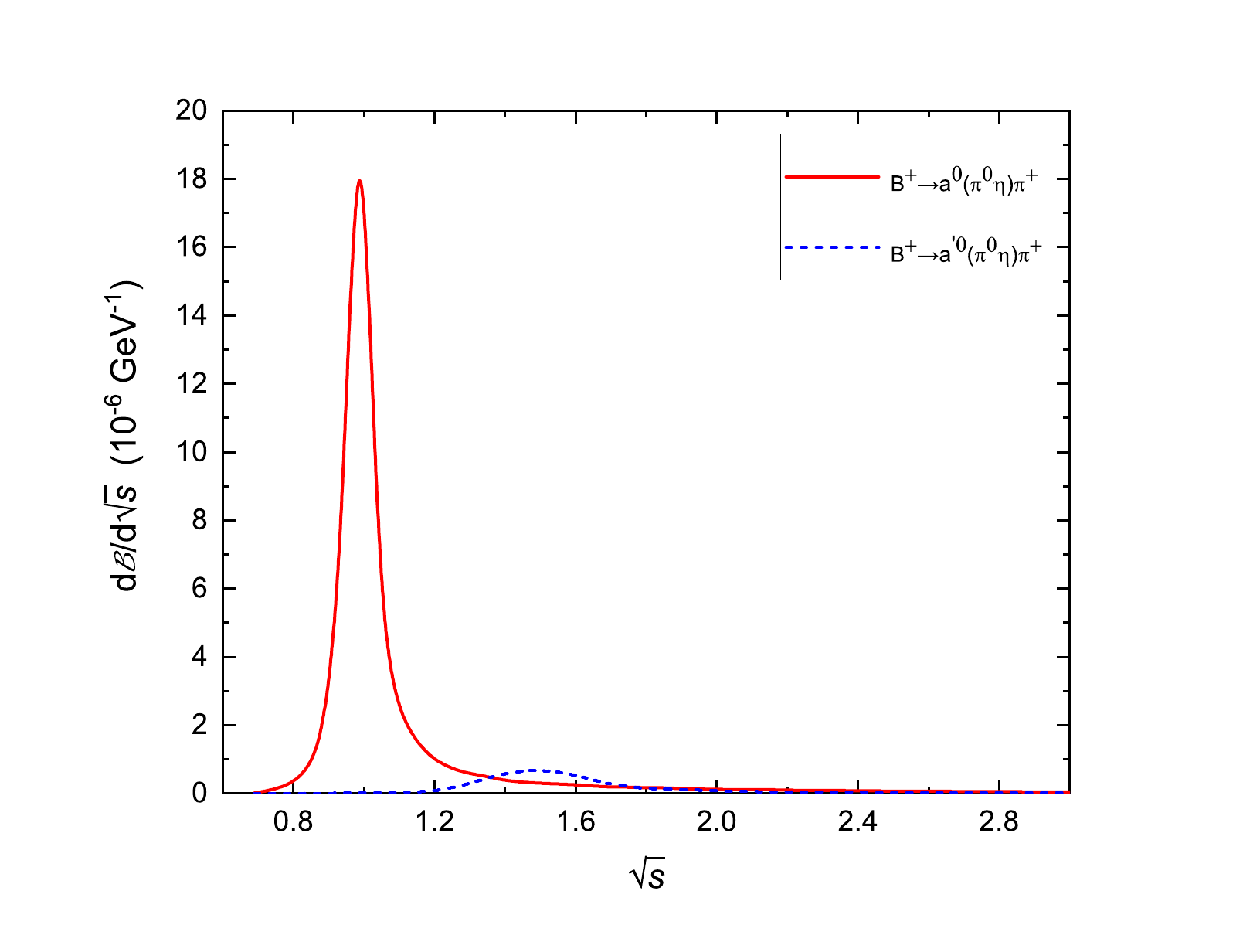}
\hspace{-1.0cm}
\includegraphics[width=0.55\textwidth]{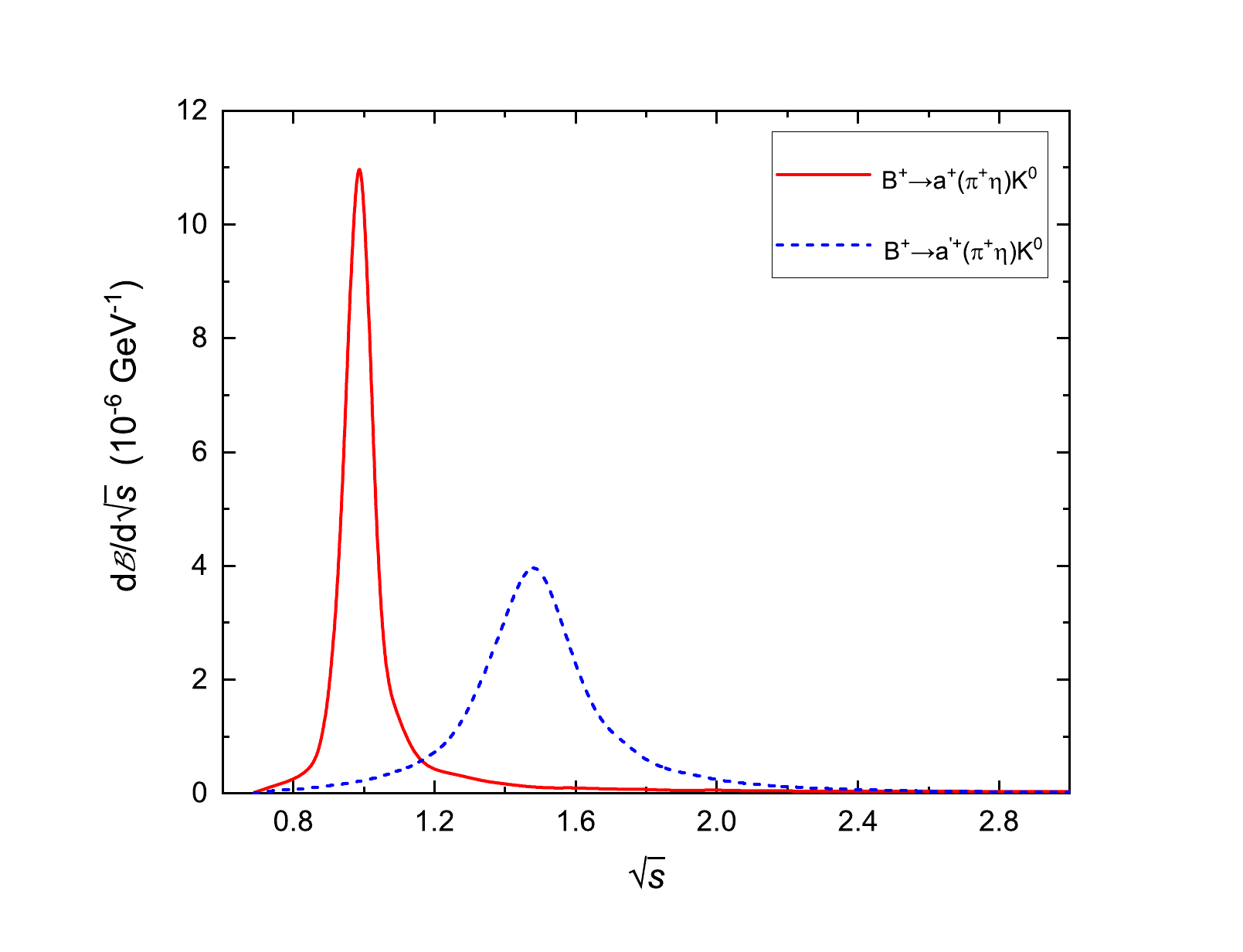} }
\vspace{-6mm}
\caption{Differential branching fractions of typical $B \to a_0^{(\prime)} \left[\to K\bar K/\pi\eta \right] h$ decays
in the first scenario of multiparticle configurations of $a_0$ mesons. }
\label{fig2}
\end{figure}

\begin{figure}[t]
\vspace{-1cm}
\centerline{\includegraphics[width=0.55\textwidth]{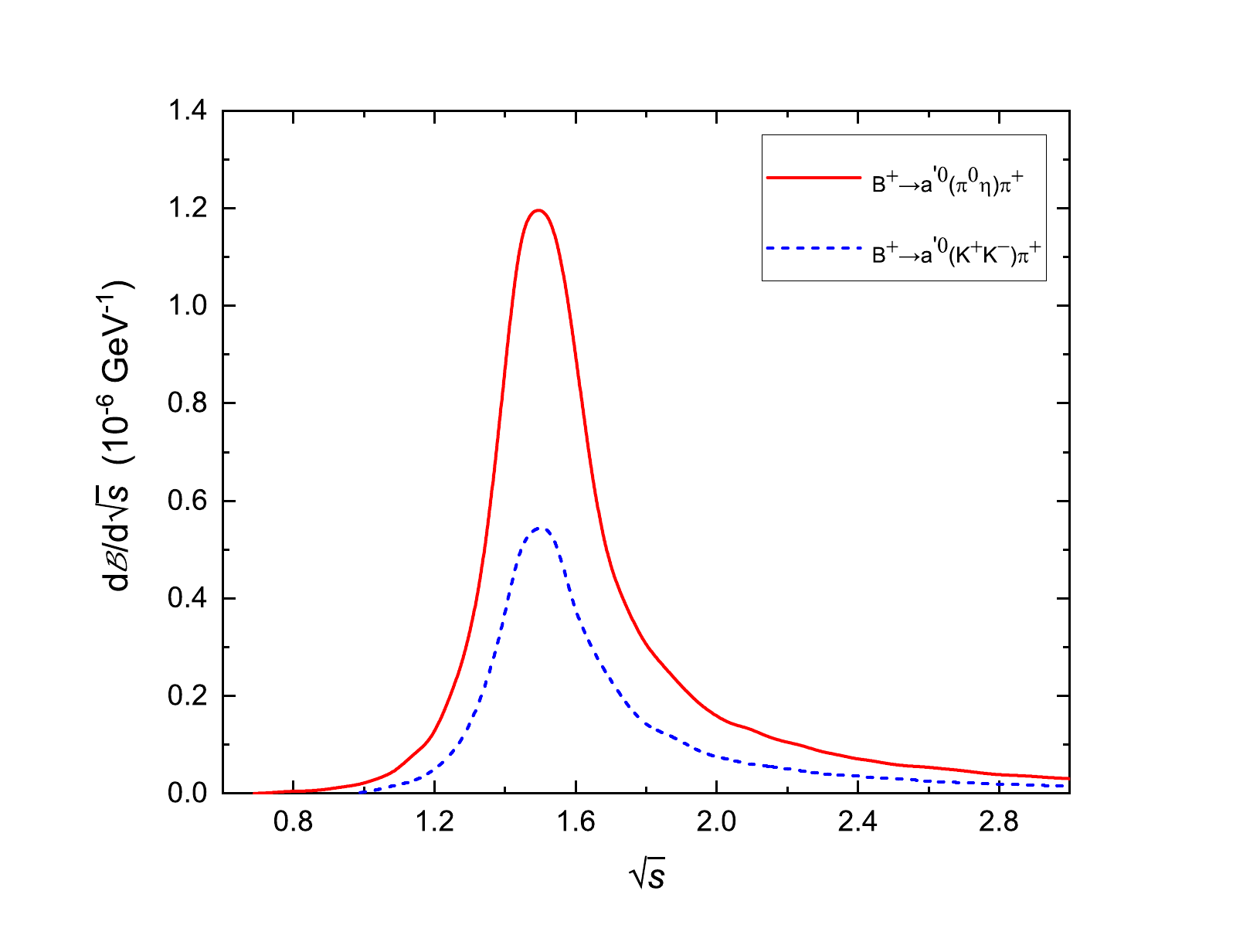}
\hspace{-1.0cm}
\includegraphics[width=0.55\textwidth]{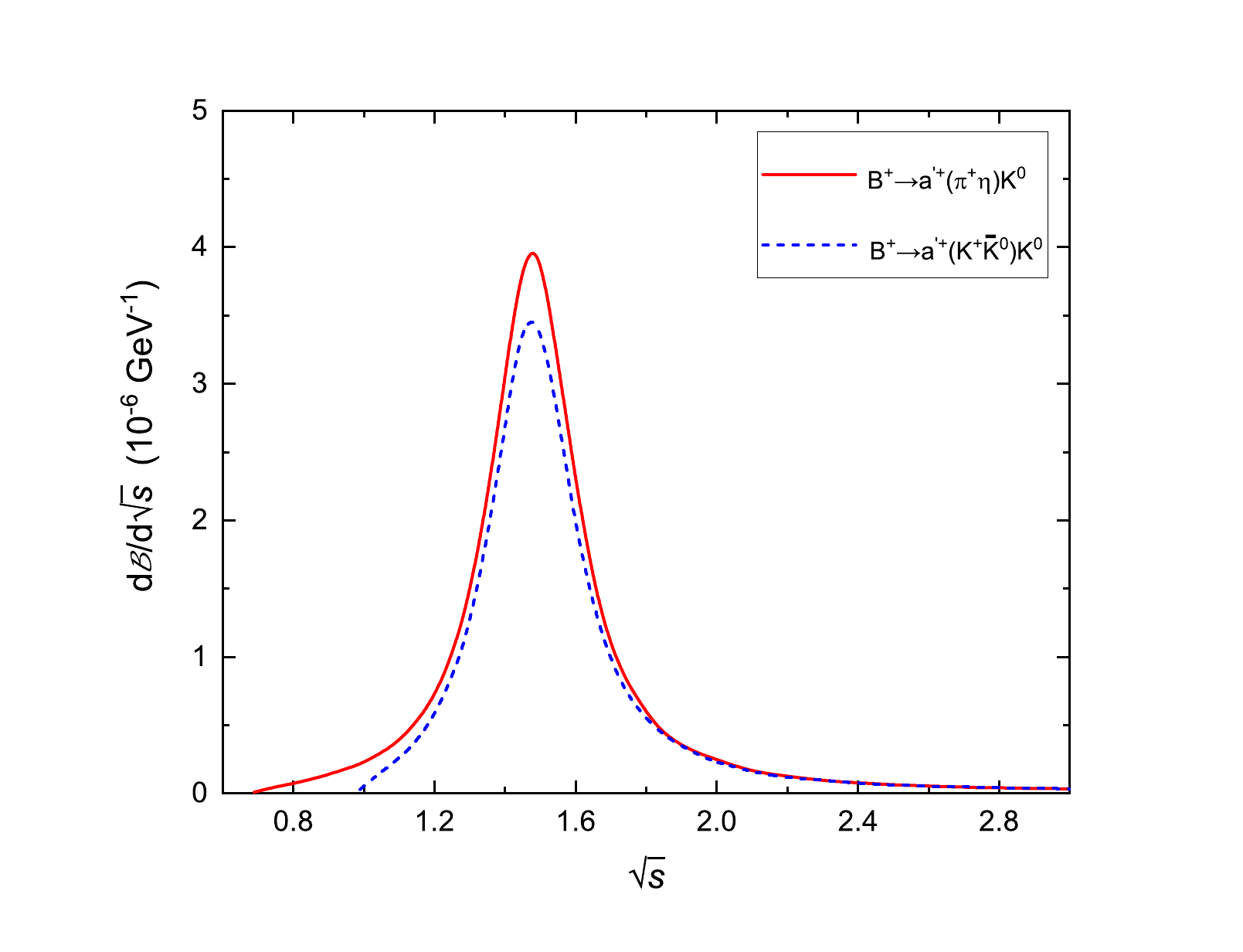} }
\vspace{-6mm}
\centerline{\includegraphics[width=0.55\textwidth]{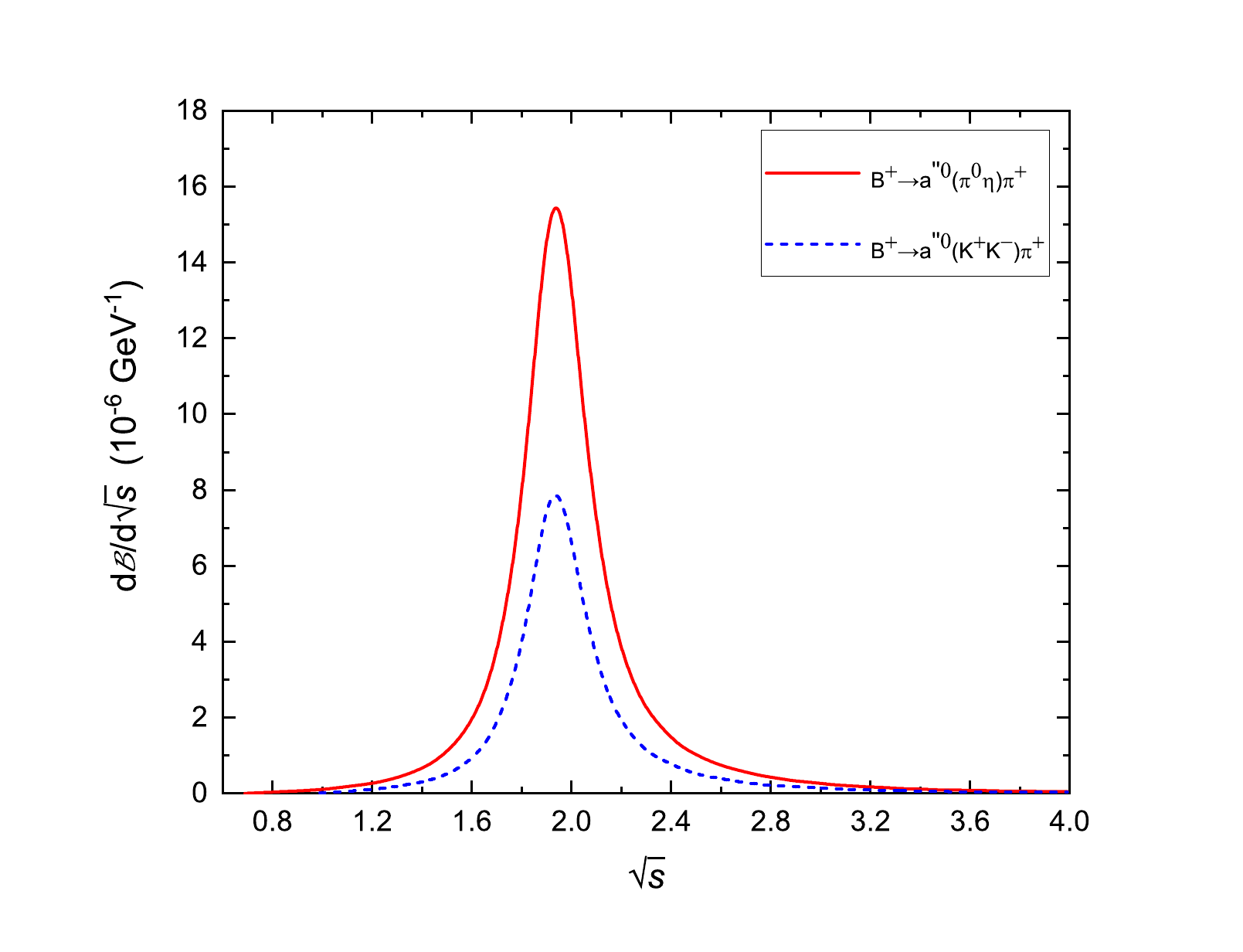}
\hspace{-1.0cm}
\includegraphics[width=0.55\textwidth]{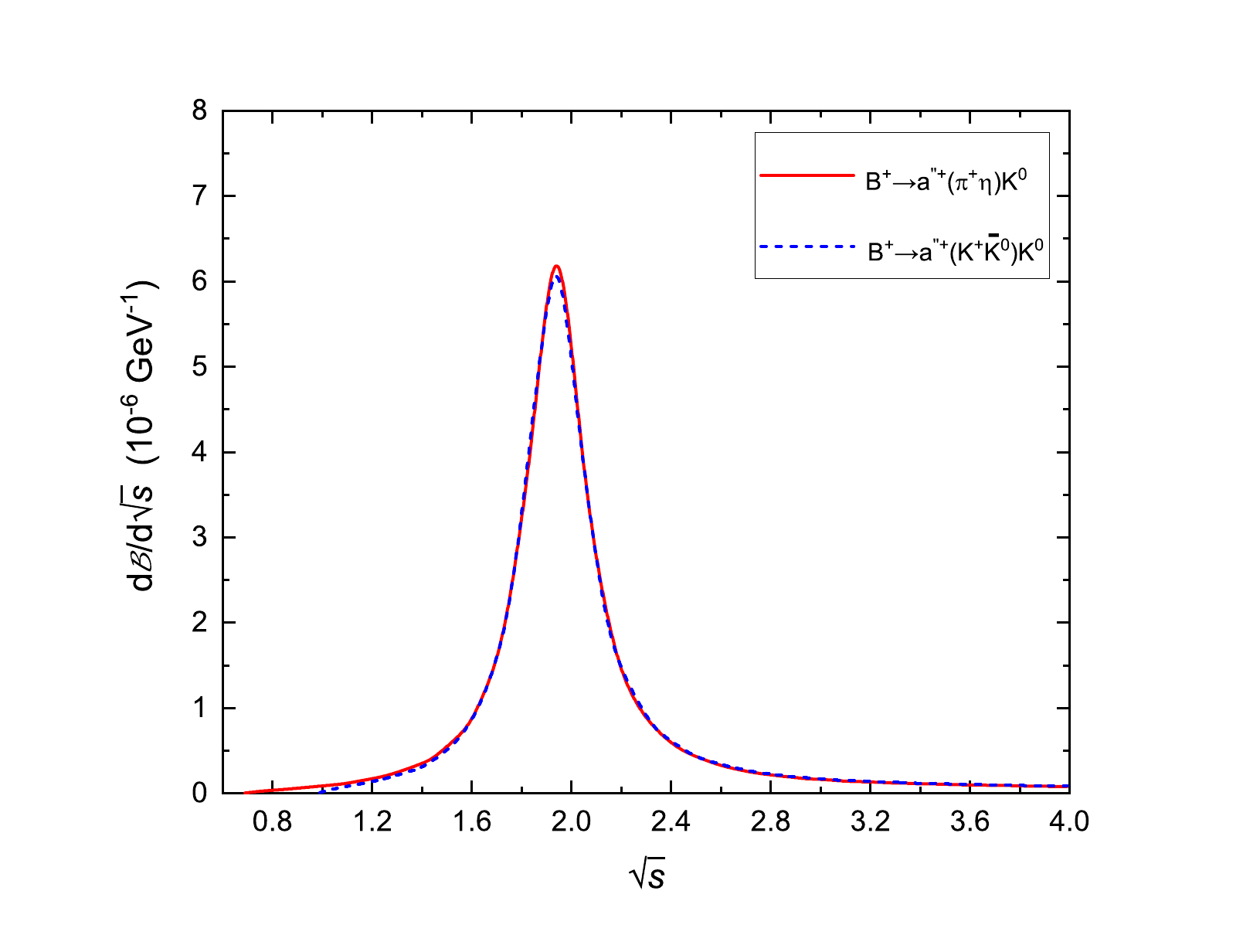} }
\vspace{-6mm}
\centerline{\includegraphics[width=0.55\textwidth]{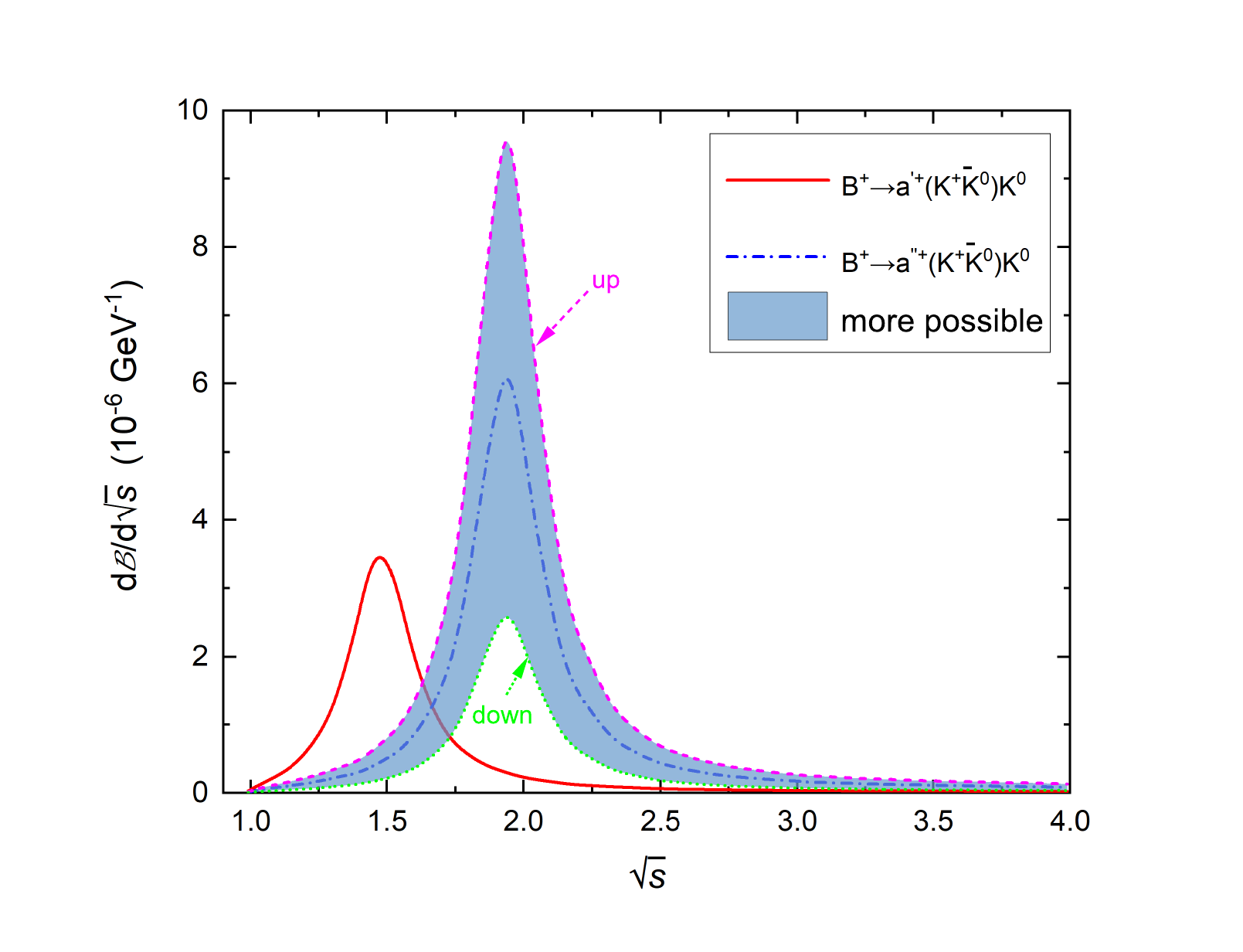}
\hspace{-1.0cm}
\includegraphics[width=0.55\textwidth]{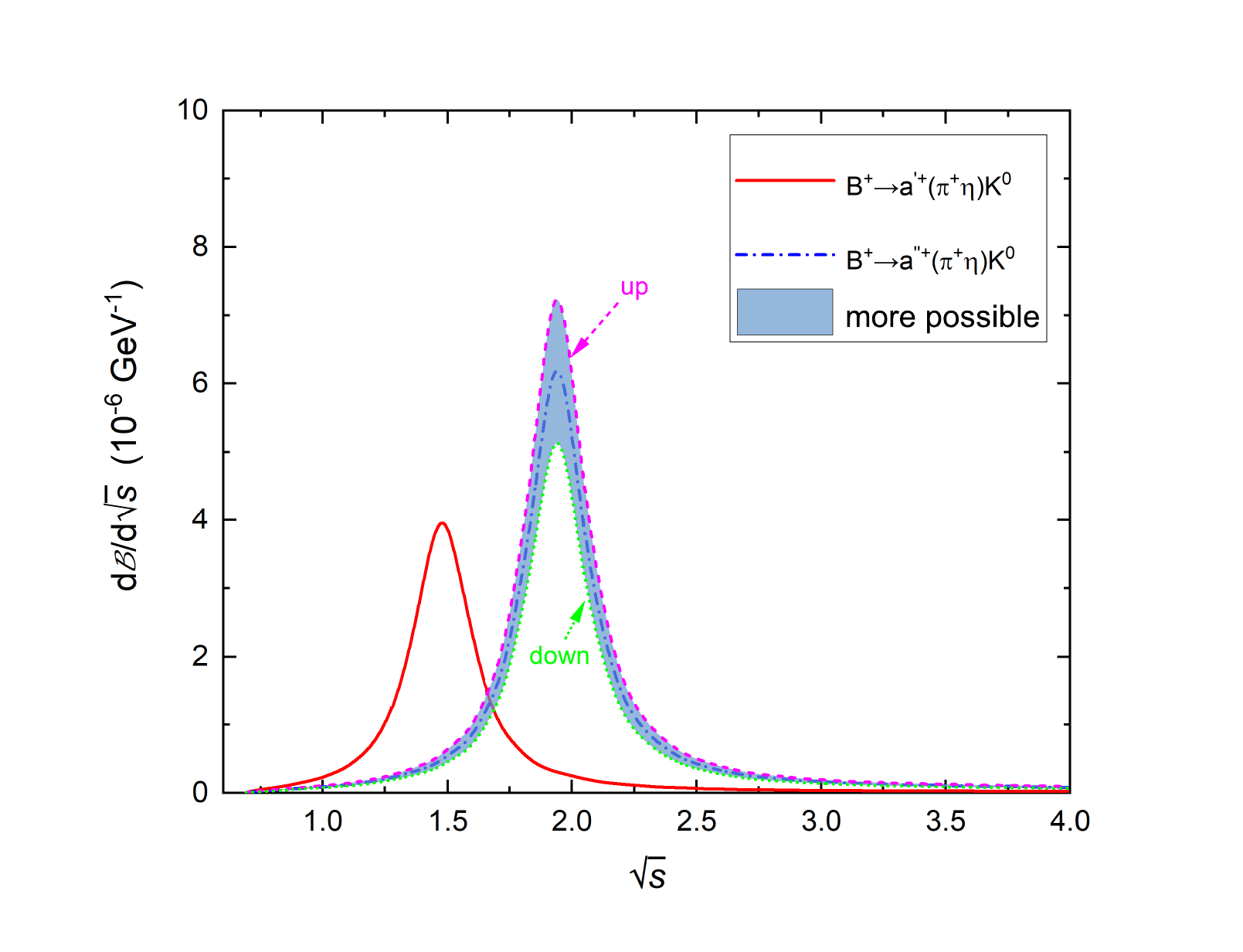} }
\vspace{-6mm}
\caption{Differential branching fractions of typical $B \to a_0^{\prime/\prime\prime} \left[\to K\bar K/\pi\eta \right] h$ decays
in the second scenario of multiparticle configurations of $a_0$ mesons.}
\label{fig3}
\end{figure}

The width effect of intermediate isovector scalar mesons is exhibited explicitly by the $K\bar K/\pi\eta$ invariant mass spectral.
In the first scenario of multiparticle configurations of $a_0$, we plot in figure \ref{fig2} for
the differential branching fractions of the typical $B \to a_0^{(\prime)} \left[\to K\bar K/\pi\eta \right] h$ decaying channels on the invariant masses,
in which the top panel shows the result of channels $B^+ \to a_0^0 [ \to K^+ K^-/\pi^0 \eta]  \pi^+$ (left) 
and $B^+ \to a_0^{+} [ \to K^+ {\bar K}^0/\pi^+ \eta] K^0$ (right) with varying the invariant mass from thresholds to $2.0 \,{\rm GeV}$,
the medium panel is the result of $B^+ \to a^{\prime 0}_0 [ \to K^+ K^-/\pi^0 \eta]  \pi^+$ (left) 
and $B^+ \to a^{\prime +}_0 [ \to K^+ {\bar K}^0/\pi^+ \eta] K^0$ (right) decays with varying the invariant mass from thresholds to $3.0 \,{\rm GeV}$,
the comparison of $a_0$ and $a_0^\prime$ contributions in $B^+ \to [\pi^0 \eta]  \pi^+$ (left) 
and $B^+ \to [\pi^+ \eta] K^0$ (right) decays is depicted in the bottom panel.
We take these typical charged channels because they carry almost all the characteristics of the relevant quasi-two-body decays:
(a) the $a_0$ contribution from $K\bar K$ mode is much smaller than it from $\pi\eta$ modes as expected by the highly phase space suppression\footnote{We multiply the result of $K\bar K$ mode by a factor of ten to show apparently for the curves.},
(b) the $a_0^\prime$ contributions from these two modes are comparable, 
we comment that the lower curves in the left plot can be compensated
by the channel $B^+ \to a^{\prime 0}_0 [ \to K^0 {\bar K}^0]  \pi^+$ which is not depicted here,
(c) in contrast to the $a_0$ contributions, 
the $a_0^\prime$ contribution is negligible in the $\left[\pi^0\eta \right]\pi^+$ channel and small in the $\left[\pi^+\eta \right] K^0$ channel,
while its contributions in the $\left[K^+K^- \right] \pi^+$ and $\left[K^+{\bar K}^0 \right] K^0$ channels 
are (much) larger than the contributions from $a_0$, this is mainly decided by the different phase spaces.
We can also see the difference between the three plots in the left panel for the channels with $h=\pi$
and the other three plots on the right panel for the channels with $h=K$, this is determined by the weak decay of relevant two-body decays
$B^+ \to a_0^{(\prime)} \pi$ and $B^+ \to a_0^{(\prime)} K$ whose invariant amplitudes are collected in the appendix \ref{sec-appx-amplitudes}.
These points support the corresponding result in tables \ref{table2} and \ref{table3}
for the partial decay branching fractions obtained by integrating the differential branching fractions over the invariant masses.

\begin{figure}[t]
\begin{center}
\vspace{-1cm}
\includegraphics[width=0.50\textwidth]{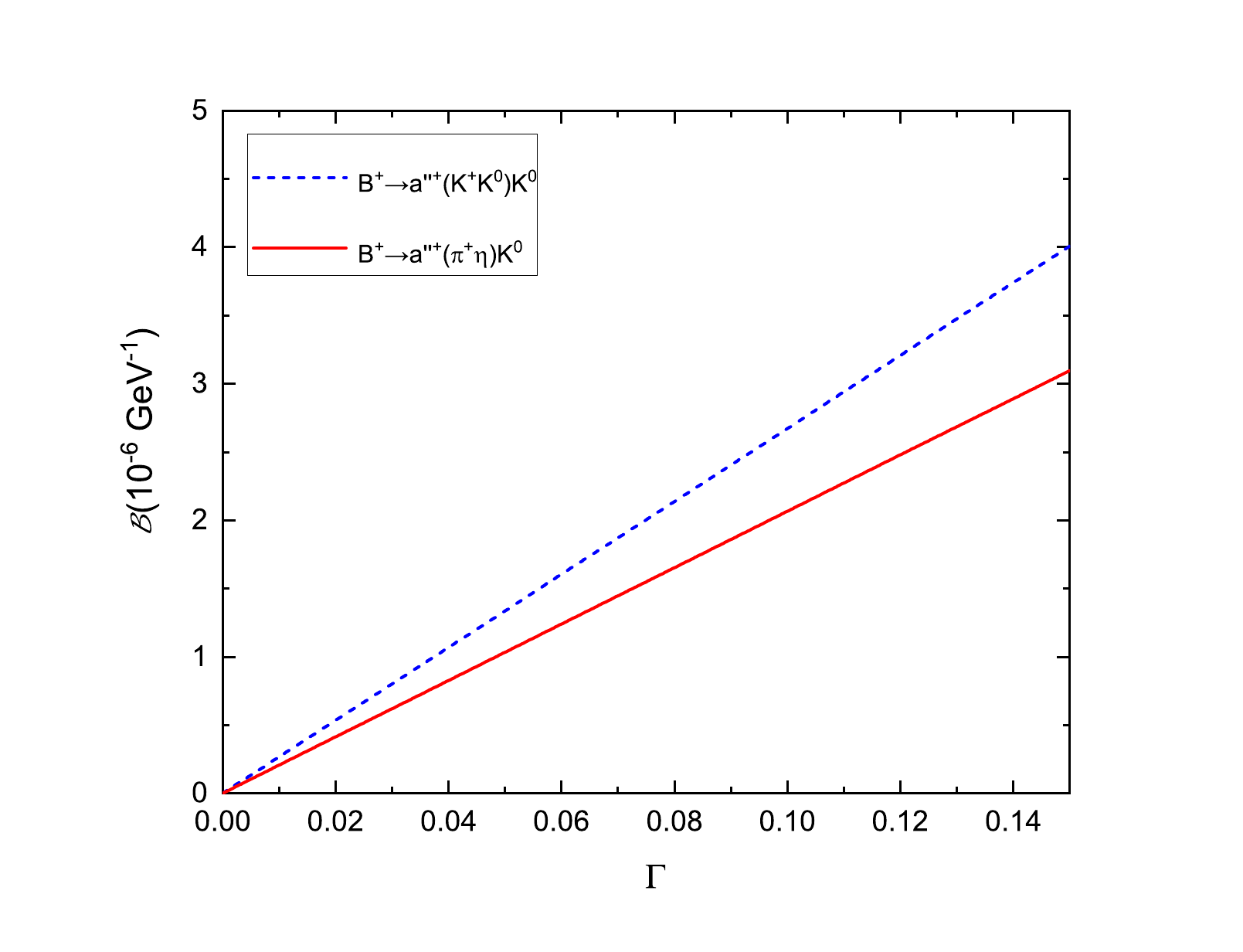}
\vspace{-0.6cm}
\caption{Evolutions of ${\cal B}(B^+ \to a_0^{\prime\prime +} \left[ \to K^+ {\bar K}^0/\pi^+\eta \right] K^0)$
on the partial widths $\Gamma_{a_0^{\prime\prime} \to K\bar K/\pi\eta}$ in the second scenario of multiparticle configurations of $a_0$ mesons.}
\label{fig4}
\end{center}
\end{figure}

We similarly plot the $a_0^{\prime}$ and $a_0^{\prime\prime}$ contributions in the typical $B \to [K\bar K/\pi\eta] h$ decay
in the second scenario of multiparticle configurations of $a_0$, as depicted in figure \ref{fig3},
where the top panel shows the result of channels
$B^+ \to a_0^{\prime 0} [ \to K^+ K^-/\pi^0 \eta]  \pi^+$ (left) and $B^+ \to a_0^{\prime +} [ \to K^+ {\bar K}^0/\pi^+ \eta] K^0$ (right)
with the invariant mass starting from the thresholds and closing up at $3.0 \,{\rm GeV}$,
the plots in medium panel is depicted for the channels
$B^+ \to a^{\prime\prime 0}_0 [ \to K^+ K^-/\pi^0 \eta]  \pi^+$ (left) and $B^+ \to a^{\prime\prime +}_0 [ \to K^+ {\bar K}^0/\pi^+ \eta] K^0$ (right)
with varying the invariant mass from thresholds to $4.0 \,{\rm GeV}$,
and the Bottom panel presents the result of channels
$B^+ \to a_0^{\prime/\prime\prime +} [ \to K^+ {\bar K}^0]  K^0$ (left) and $B^+ \to a_0^{\prime/\prime\prime +} [ \to \pi^+ \eta] K^0$ (right).
We can easily get that
(a) the contributions from $a_0^\prime$ in the channels $B^+ \to \left[\pi^0 \eta/K^+K^- \right] \pi^+$ 
and $B^+ \to \left[\pi^+\eta/K^+{\bar K}^0 \right] K^0$
in the second scenario of multiparticle configurations of $a_0$ are very close to that obtained in the first scenario,
we would like to mark again that the neutral $B$ meson decaying channels $B^0 \to [K^\pm K^0/\pi^\pm \eta] \pi^{\mp}$,
even though have the similar shapes, have apparent different predictions in magnitude in these two scenarios, 
(b) the contributions from $a_0^{\prime\prime}$ are larger than that from $a_0^\prime$,
in the $\left[ \pi^0 \eta/K^+K^- \right] \pi^+$ channels even larger by about a order,
this is an impressive result but not surprise if we look at the twist 2 LCDAs in Eq. (\ref{def-wavefun-twist2}) and the relevant parameters,
and we looking forward for the experiment check,
(c) the $a_0^{\prime\prime}$ contributions in the channels $\left[\pi^+\eta \right] K^0$ and $\left[K^+{\bar K}^0 \right] K^0$ are almost overlap
because the $a_0^{\prime\prime}$ is far away from the $K\bar K$ and $\pi\eta$ thresholds,
as we can also find in the channels $\left[\pi^0\eta \right] \pi^+$ and $\left[K\bar K \right] \pi^+$ if we consider both the $K^+K^-$ and $K^0{\bar K}^0$ contributions,
(d) the partial widths of $a_0^{\prime\prime} \to K\bar K/\pi\eta$ effect significantly for the result of the quasi-two-body,
we plot the varying band in the bottom panel by taking the result $\Gamma_{a_0^{\prime\prime}}\to K\bar K = 94 \pm 54 \, {\rm MeV}$ and
$\Gamma_{a_0^{\prime\prime}} \to \pi\eta = 94 \pm 16 \, {\rm MeV}$ obtained from the eLSM model \cite{Parganlija:2016yxq}.
We depict in figure \ref{fig4} the dependence of the branching fractions of $B^+ \to a_0^{\prime\prime +} \left[ \to K^+ {\bar K}^0/\pi^+\eta \right] K^0$
on the partial widths $\Gamma_{a_0^{\prime\prime} \to K\bar K/\pi \eta}$ with considering the largest uncertainties in Eq. (\ref{eq:partial-width}).
It is shown that the width effect of $a_0^{\prime\prime}$ in the relevant quasi-two-body $B$ decays is linear,
so we suggest these channels in $B$ decays to determine the partial widths $\Gamma_{a_0^{\prime\prime} \to K\bar K/\pi \eta}$.

\section{Conclusion}\label{sec-conclusion}

Motivated by the discrepancy between the experimental measurements of three-body $B \to a_0(980) \left[ \to \pi\eta \right] K$ decays
and the theoretical predictions of two-body $B \to a_0(980)K$ decays,
we study the contributions from $a_0$ in the three-body $B \to \left[\pi\eta \right](\left[K{\bar K} \right]) h$ decays in the framework of PQCD approach,
where the width effects of the intermediated isovector scalar mesons $a_0$ are demonstrated in detail,
this is also the first systematical study of the width effect in $B \to a_0$ decays.
In the face of controversy for the multipaticle configurations of $a_0(980)$, particularly in the $B$ decays,
we consider two scenarios where the first one states that $a_0(980)$ is the lowest-lying $q{\bar q}$ state,
and the second one says that the lowest-lying $q{\bar q}$ state is $a_0(1450)$
while $a_0(980)$ is a compact tetraquark state or $K{\bar K}$ bound state.

We find that the width effect from intermediate $a_0$ states is significant in the relevant quasi-two-body decaying channels, 
with which we extract the branching fractions of corresponding two-body decays under narrow width approximation, 
showing a large difference to the previous direct two-body calculation under the static $a_0(980)$ assumption. 
Our calculations show that the $a_0(980)$ as the lowest-lying $q{\bar q}$ state
can not be ruled out in $B$ decays within acceptable limits with the current measurements. 
To examine the nature of $a_0$ state in $B$ decays, we suggest several channels for the future experiments.   
The first candidate is the $B \to a_0^- [\to \pi^- \eta] \pi^+$ mode with the largest branching fraction from the calculation under the first scenario,  
the second ones are the $B^0 \to a_0^\pm(1450) \left[\to K^\pm{\bar K}^0/\pi^\pm \pi^0 \right] \pi^\mp$ modes, 
whose branching fractions obtained in the first scenario is about three times smaller in magnitude than that obtained in the second scenario, 
the last, but not the least, is the partial widths ($\Gamma_{a_0(1950) \to K{\bar K}/\pi\eta}$) dependence of the partial branching fractions 
of $B \to a_0(1950) \left[ K{\bar K}/\pi\eta\right] h$ modes, 
this dependence is shown in the linear behaviour and could be examined by the future data.
As a byproduct, we present $a_0$ mesons contributions in the ${\rm CKM}$ suppressed $B_s$ decays,
which seems more harder for the near future experiments.

\section{Acknowledgments}
We would like to thank Wen-fei Wang for proposing this project to us, and to Hai-yang Cheng for the fruitful discussion.
This work is supported by the National Science Foundation of China (NSFC) under the Nos. 11805060, 11975112,11947011 
and the Joint Large Scale Scientific Facility Funds of the NSFC and CAS under Contract No. U1932110. 
{\bf SC} is also supported by the Natural Science Foundation of Hunan Province, China (Grant No. 2020JJ4160), 
{\bf AJM} is also supported by the Natural Science Foundation of Jiangsu Province, China (Grant No. BK20191010) 
and the Scientific Research Foundation of Nanjing Institute of Technology (Grant No. YKJ201854).

\appendix

\section{Probing $a_0$ mesons in the quasi-two-body $B_s$ decays}\label{sec-Bs}

We also predict the contributions from isovector scalar mesons in the ${\rm CKM}$ suppressed $B_s$ decays under, 
as presented in table \ref{table6} and table \ref{table7} under scenario I and II, respectively, 
the channel $(B_s^0 \to a_0^- \left[ \pi^- \eta \right] K^+)$ with the predicted branching fraction 
$(0.75^{+0.22+0.13}_{-0.14-0.12} ) \times 10^{-6}$ is the most possible available at the near future experiments. 

\begin{table}[hbt]
\vspace{-2mm}
\caption{The same as table \ref{table2}, but for the $B^0_s \to a^{(\prime)}_0 \left[ \to K\bar K/ \pi\eta \right] h$ decays.}
\begin{center}
\begin{tabular}{l|c|c|r}
\toprule
{\rm Decay modes} & \; {\rm Quasi-two-body} \; & \; {\rm narrow approx.} \; & \quad {\rm CPV} \quad\quad \non
\hline
$B^0_s \to a_0^+ \left[\to K^+{\bar K}^0\right] \pi^-$ &$0.03^{+0.00+0.01}_{-0.00-0.01}$ &&$-8.1^{+6.0+4.8}_{-2.7-7.1}$ \non
$\quad\;\, \to a_0^+ \left[\to \pi^+ \eta\right] \pi^-$ &$0.17^{+0.00+0.06}_{-0.00-0.04}$ &$0.25^{+0.00+0.09}_{-0.01-0.05}$ &$-11.7^{+2.7+1.5}_{-9.3-5.7}$ \non
$B^0_s \to a_0^0 \left[\to K^-K^+\right] \pi^0$ &$0.04^{+0.00+0.01}_{-0.01-0.01}$ & &$16.5^{+2.5+8.0}_{-0.3-3.5}$ \non
$\quad\;\, \to a_0^0 \left[\to \pi^0 \eta\right] \pi^0$ &$0.49^{+0.03+0.16}_{-0.06-0.11}$ &$0.70^{+0.04+0.24}_{-0.09-0.14}$ &$22.7^{+1.2+2.6}_{-1.5-3.4}$ \non
$B^0_s \to a_0^- \left[\to K^-K^0\right] \pi^+$ &$0.03^{+0.01+0.00}_{-0.01-0.00}$ &&$22.1^{+6.5+12.9}_{-3.1-10.8}$ \non
$\quad\;\, \to a_0^- \left[\to \pi^- \eta\right] \pi^+$ &$0.14^{+0.05+0.09}_{-0.02-0.02}$ &$0.20^{+0.08+0.04}_{-0.03-0.03}$ &$44.4^{+4.2+2.9}_{-9.5-8.4}$ \non
$B^0_s \to a_0^0 \left[\to K^-K^+\right] K^0$ &$0.16^{+0.05+0.02}_{-0.03-0.01}$ &&$54.8^{+0.1+6.4}_{-5.3-7.5}$ \non
$\quad\;\, \to a_0^0 \left[\to \pi^0 \eta\right] K^0$ &$0.74^{+0.19+0.09}_{-0.15-0.07}$ &$1.05^{+0.28+0.12}_{-0.21-0.10}$ &$61.4^{+1.3+5.9}_{-1.6-7.5}$ \non
$B^0_s \to a_0^- \left[\to K^-K^0\right] K^+$ &$0.07^{+0.02+0.01}_{-0.01-0.01}$ &&$81.5^{+4.8+1.5}_{-8.8-3.5}$ \non
$\quad\;\, \to a_0^- \left[\to \pi^- \eta\right] K^+$ &$0.75^{+0.22+0.14}_{-0.13-0.12}$ &$1.06^{+0.32+0.20}_{-0.19-0.18}$ &$77.8^{+1.6+5.6}_{-9.1-3.1}$ \non
\hline
$B^0_s \to a_0^{\prime +} \left[\to K^+{\bar K}^0\right] \pi^-$ &$0.09^{+0.02+0.03}_{-0.02-0.02}$ &$1.05^{+0.29+0.33}_{-0.21-0.28}$ &$-10.1^{+0.2+1.8}_{-0.8-0.3}$ \non
$\quad\;\, \to a_0^{\prime +} \left[\to \pi^+ \eta\right] \pi^-$ &$0.10^{+0.03+0.03}_{-0.02-0.03}$ &$1.08^{+0.31+0.33}_{-0.22-0.31}$ &$-8.3^{+1.8+1.8}_{-3.0-3.9}$ \non
$B^0_s \to a_0^{\prime 0} \left[\to K^-K^+\right] \pi^0$ &$0.13^{+0.03+0.03}_{-0.02-0.03}$ &$3.07^{+0.61+0.84}_{-0.56-0.71}$ &$19.7^{+1.9+2.0}_{-0.4-3.8}$ \non
$\quad\;\, \to a_0^{\prime 0} \left[\to \pi^0 \eta\right] \pi^0$ &$0.30^{+0.06+0.08}_{-0.06-0.07}$ &$3.20^{+0.68+0.84}_{-0.61-0.71}$ &$20.8^{+1.2+2.0}_{-0.2-3.9}$ \non
$B^0_s \to a_0^{\prime -} \left[\to K^-K^0\right] \pi^+$ &$0.05^{+0.01+0.02}_{-0.01-0.01}$ &$0.67^{+0.16+0.22}_{-0.12-0.18}$ &$53.9^{+0.5+1.5}_{-3.8-3.7}$ \non
$\quad\;\, \to a_0^{\prime -} \left[\to \pi^- \eta\right] \pi^+$ &$0.06^{+0.02+0.02}_{-0.01-0.02}$ &$0.69^{+0.18+0.21}_{-0.13-0.18}$ &$55.4^{+0.5+2.7}_{-2.0-2.5}$ \non
$B^0_s \to a_0^{\prime 0} \left[\to K^-K^+\right] K^0$ &$0.07^{+0.01+0.02}_{-0.01-0.02}$ &$0.88^{+0.12+0.29}_{-0.09-0.29}$ &$-6.6^{+0.9+9.2}_{-5.0-12.1}$ \non
$\quad\;\, \to a_0^{\prime 0} \left[\to \pi^0 \eta\right] K^0$ &$0.08^{+0.01+0.03}_{-0.01-0.02}$ &$0.89^{+0.13+0.30}_{-0.08-0.28}$ &$-8.8^{+2.2+10.4}_{-5.9-12.1}$ \non
$B^0_s \to a_0^{\prime -} \left[\to K^-K^0\right] K^+$ &$0.03^{+0.01+0.01}_{-0.00-0.01}$ &$0.64^{+0.11+0.25}_{-0.04-0.12}$ &$-28.3^{+2.7+10.1}_{-9.1-8.6}$ \non
$\quad\;\, \to a_0^{\prime -} \left[\to \pi^- \eta\right] K^+$ &$0.06^{+0.01+0.02}_{-0.00-0.01}$ &$0.66^{+0.11+0.23}_{-0.04-0.13}$ &$-27.8^{+1.4+6.1}_{-6.1-8.6}$ \non
\toprule
\end{tabular}
\end{center}
\label{table6}
\end{table}

\begin{table}[t]
\vspace{-2mm}
\caption{The same as table \ref{table4}, but for the $B^0_s \to a^{\prime/\prime\prime}_0 \left[ \to K\bar K/ \pi\eta \right] h$ decays.}
\begin{center}
\begin{tabular}{l|c|c|r}
\toprule
{\rm Decay modes} & \; {\rm Quasi-two-body} \; & \; {\rm narrow approx.} \; & \quad {\rm CPV} \quad\quad \non
\hline
$B^0_s \to a_0^{\prime +} \left[\to K^+{\bar K}^0\right] \pi^-$ &$0.08^{+0.02+0.03}_{-0.01-0.03}$ &$1.02^{+0.20+0.39}_{-0.18-0.35}$ &$-0.9^{+1.0+1.0}_{-0.2-0.7}$ \non
$\quad\;\, \to a_0^{\prime +} \left[\to \pi^+ \eta\right] \pi^-$ &$0.10^{+0.02+0.04}_{-0.02-0.03}$ &$1.05^{+0.20+0.40}_{-0.19-0.36}$ &$-0.9^{+1.6+0.3}_{-0.3-0.3}$ \non
$B^0_s \to a_0^{\prime 0} \left[\to K^-K^+\right] \pi^0$ &$0.11^{+0.02+0.04}_{-0.02-0.04}$ &$2.73^{+0.58+1.11}_{-0.49-0.92}$ &$17.9^{+0.3+1.5}_{-1.3-2.3}$ \non
$\quad\;\, \to a_0^{\prime 0} \left[\to \pi^0 \eta\right] \pi^0$ &$0.26^{+0.06+0.10}_{-0.05-0.08}$ &$2.79^{+0.50+1.14}_{-0.50-0.92}$ &$16.2^{+0.6+0.8}_{-0.1-1.0}$ \non
$B^0_s \to a_0^{\prime -} \left[\to K^-K^0\right] \pi^+$ &$0.03^{+0.01+0.02}_{-0.01-0.01}$ &$0.36^{+0.07+0.14}_{-0.06-0.17}$ &$26.8^{+4.8+4.8}_{-8.5-6.5}$ \non
$\quad\;\, \to a_0^{\prime -} \left[\to \pi^- \eta\right] \pi^+$ &$0.03^{+0.01+0.02}_{-0.01-0.02}$ &$0.36^{+0.09+0.25}_{-0.05-0.16}$ &$22.4^{+6.7+1.5}_{-2.2-2.2}$ \non
$B^0_s \to a_0^{\prime 0} \left[\to K^-K^+\right] K^0$ &$0.15^{+0.04+0.04}_{-0.02-0.03}$ &$1.88^{+0.44+0.39}_{-0.28-0.33}$ &$22.6^{+2.2+5.1}_{-1.6-3.5}$ \non
$\quad\;\, \to a_0^{\prime 0} \left[\to \pi^0 \eta\right] K^0$ &$0.17^{+0.04+0.04}_{-0.03-0.03}$ &$1.88^{+0.44+0.40}_{-0.27-0.31}$ &$23.4^{+2.2+4.4}_{-1.9-4.2}$ \non
$B^0_s \to a_0^{\prime -} \left[\to K^-K^0\right] K^+$ &$0.04^{+0.01+0.01}_{-0.01-0.01}$ &$1.07^{+0.27+0.23}_{-0.16-0.19}$ &$57.1^{+0.1+6.1}_{-1.4-5.4}$ \non
$\quad\;\, \to a_0^{\prime -} \left[\to \pi^- \eta\right] K^+$ &$0.10^{+0.02+0.03}_{-0.01-0.01}$ &$1.09^{+0.25+0.26}_{-0.16-0.19}$ &$57.6^{+0.2+7.1}_{-0.9-4.8}$ \non
\hline
$B^0_s \to a_0^{\prime\prime +} \left[\to K^+{\bar K}^0\right] \pi^-$ &$0.02^{+0.01+0.06}_{-0.01-0.02}\pm 0.01$ &$0.07^{+0.03+0.17}_{-0.01-0.06}$ &$50.9^{+6.0+13.3}_{-9.3-11.0}$ \non
$\quad\;\, \to a_0^{\prime\prime +} \left[\to \pi^+ \eta\right] \pi^-$ &$0.03^{+0.01+0.06}_{-0.01-0.03}\pm 0.01$ &$0.08^{+0.02+0.17}_{-0.02-0.07}$ &$48.5^{+3.7+8.6}_{-9.6-10.5}$ \non
$B^0_s \to a_0^{\prime\prime 0} \left[\to K^-K^+\right] \pi^0$ &$0.03^{+0.01+0.08}_{-0.0-0.04}\pm 0.02$ &$0.20^{+0.08+0.48}_{-0.01-0.17}$ &$37.0^{+3.8+6.0}_{-0.9-11.1}$ \non
$\quad\;\, \to a_0^{\prime\prime 0} \left[\to \pi^0 \eta\right] \pi^0$ &$0.07^{+0.02+0.16}_{-0.01-0.07}\pm 0.01$ &$0.22^{+0.05+0.48}_{-0.03-0.19}$ &$42.7^{+4.1+13.2}_{-3.7-5.8}$ \non
$B^0_s \to a_0^{\prime\prime -} \left[\to K^-K^0\right] \pi^+$ &$0.10^{+0.01+0.08}_{-0.01-0.05}\pm 0.06$ &$0.30^{+0.03+0.24}_{-0.04-0.17}$ &$35.3^{+2.0+13.7}_{-0.4-9.2}$ \non
$\quad\;\, \to a_0^{\prime\prime -} \left[\to \pi^- \eta\right] \pi^+$ &$0.10^{+0.01+0.08}_{-0.02-0.06}\pm 0.02$ &$0.31^{+0.03+0.23}_{-0.05-0.18}$ &$31.0^{+3.4+8.7}_{-0.5-7.5}$ \non
$B^0_s \to a_0^{\prime\prime 0} \left[\to K^-K^+\right] K^0$ &$0.69^{+0.24+0.24}_{-0.16-0.18}\pm 0.39$ &$2.06^{+0.71+0.74}_{-0.48-0.56}$ &$-3.6^{+3.4+7.5}_{-3.5-8.8}$ \non
$\quad\;\, \to a_0^{\prime\prime 0} \left[\to \pi^0 \eta\right] K^0$ &$0.69^{+0.24+0.25}_{-0.16-0.19}\pm 0.12$ &$2.06^{+0.71+0.74}_{-0.48-0.57}$ &$-3.8^{+3.4+7.5}_{-3.6-9.3}$ \non
$B^0_s \to a_0^{\prime\prime -} \left[\to K^-K^0\right] K^+$ &$0.20^{+0.07+0.08}_{-0.05-0.05}\pm 0.11$ &$1.19^{+0.41+0.47}_{-0.28-0.33}$ &$-6.6^{+8.8+10.6}_{-8.9-10.9}$ \non
$\quad\;\, \to a_0^{\prime\prime -} \left[\to \pi^- \eta \right] K^+$ &$0.40^{+0.14+0.16}_{-0.09-0.10}\pm 0.07$ &$1.18^{+0.43+0.50}_{-0.27-0.32}$ &$-7.6^{+9.9+9.5}_{-7.7-10.3}$ \non
\toprule
\end{tabular}
\end{center}
\label{table7}
\end{table}

\section{Decay amplitudes}\label{sec-appx-amplitudes}

In this section, we list the Lorentz invariant decay amplitude ${\mathcal A}$ for the considered quasi-two-body decay in the PQCD approach.
\beq
{\cal A}(B^+ \to a_0^+\pi^0) &=& \frac{G_F} {2}
 \big\{V_{ub}^*V_{ud}[(a_1(F^{LL}_{Th}+F^{LL}_{Ah}-F^{LL}_{Aa_0})+a_2F^{LL}_{Ta_0}+C_1(M^{LL}_{Th}+M^{LL}_{Ah}-M^{LL}_{Aa_0})\nonumber\\
&+&C_2 M^{LL}_{Ta_0}]-V_{tb}^*V_{td}[(-a_4+\frac{5 C_9}{3}+C_{10}-\frac{3 a_7}{2})F^{LL}_{Ta_0}-(a_6-\frac{a_8}{2})F^{SP}_{Ta_0}\nonumber\\
&+&(\frac{C_9+3 C_{10}}{2}-C_3)M^{LL}_{Ta_0}-(C_5-\frac{C_7}{2})M^{LR}_{Ta_0}+\frac{3 C_8}{2}M^{SP}_{Ta_0}\nonumber\\
&+&(a_4+a_{10})(F^{LL}_{Th}+F^{LL}_{Ah}-F^{LL}_{Aa_0})+(a_6+a_8)(F^{SP}_{Th}+F^{SP}_{Ah}-F^{SP}_{Aa_0})\nonumber\\
&+&(C_3+C_9)(M^{LL}_{Th}+M^{LL}_{Ah}-M^{LL}_{Aa_0})+(C_5+C_7)(M^{LR}_{Th}+M^{LR}_{Ah}-M^{LR}_{Aa_0})]\big\} \;,
\label{amp1} \\
 {\cal A}(B^+ \to a_0^0\pi^+) &=&
 \frac{G_F} {2}\big\{V_{ub}^*V_{ud}[a_1(F^{LL}_{Ta_0}+F^{LL}_{Aa_0}-F^{LL}_{Ah})+a_2F^{LL}_{Th}+C_1(M^{LL}_{Ta_0}+M^{LL}_{Aa_0}-M^{LL}_{Ah})\nonumber\\
&+&C_2 M^{LL}_{Th}]-V_{tb}^*V_{td}[(a_4+a_{10})(F^{LL}_{Ta_0}+F^{LL}_{Aa_0}-F^{LL}_{Ah})-(a_6-\frac{a_8}{2})F^{SP}_{Th}\nonumber\\
&+&(a_6+a_8)(F^{SP}_{Ta_0}+F^{SP}_{Aa_0}-F^{SP}_{Ah})+(C_3+C_9)(M^{LL}_{Ta_0}+M^{LL}_{Aa_0}-M^{LL}_{Ah})\nonumber\\
&+&(C_5+C_7)(M^{LR}_{Ta_0}+M^{LR}_{Aa_0}-M^{LR}_{Th})+(\frac{5}{3}C_9+C_{10}+\frac{3 a_7}{2}-a_4)F^{LL}_{Th}\nonumber\\
&+&(\frac{C_9+3 C_{10}}{2}-C_3)M^{LL}_{Th}-(C_5-\frac{C_7}{2})M^{LR}_{Th}+\frac{3 C_8}{2}M^{SP}_{Th}]\big\} \;,
\label{amp2}\\
 {\cal A}(B^+ \to a_0^+K^0) &=&
  \frac{G_F} {\sqrt{2}}\big\{V_{ub}^*V_{us}[a_1F^{LL}_{Aa_0}+C_1 M^{LL}_{Aa_0}]-V_{tb}^*V_{ts}[(a_4-\frac{a_{10}}{2})F^{LL}_{Ta_0}\nonumber\\
&+&(a_6-\frac{a_8}{2})F^{SP}_{Ta_0}+(C_3-\frac{C_9}{2})M^{LL}_{Ta_0}+(C_5-\frac{C_7}{2})M^{LR}_{Ta_0}+(a_4+a_{10})F^{LL}_{Aa_0}\nonumber\\
&+&(C_3+C_9)M^{LL}_{Aa_0}+(a_6+a_8)F^{SP}_{Aa_0}+(C_5+C_7)M^{LR}_{Aa_0}]\big\} \;,
\label{amp3}\\
 {\cal A}(B^+ \to a_0^0K^+) &=& \frac{G_F} {2} \big\{V_{ub}^*V_{us}[a_1(F^{LL}_{Ta_0}+F^{LL}_{Aa_0})+a_2 F^{LL}_{Th}+C_1(M^{LL}_{Ta_0}+M^{LL}_{Aa_0})\nonumber\\
&+&C_2M^{LL}_{Th}]-V_{tb}^*V_{ts}[(a_4+a_{10})(F^{LL}_{Ta_0}+F^{LL}_{Aa_0})+(a_6+a_8)(F^{SP}_{Ta_0}+F^{SP}_{Aa_0})\nonumber\\
&+&(C_3+C_9)(M^{LL}_{Ta_0}+M^{LL}_{Aa_0})+(C_5+C_7)(M^{LR}_{Ta_0}+M^{LR}_{Aa_0})\nonumber\\
&+&\frac{3}{2}(a_7+a_9)F^{LL}_{Th}+\frac{3 C_{10} }{2} M^{LL}_{Th} +\frac{3 C_8}{2} M^{SP}_{Th}]\big\} \;,
\label{amp4} \\
 {\cal A}(B^0 \to a_0^+\pi^-) &=& \frac{G_F} {\sqrt{2}}
 \big\{V_{ub}^*V_{ud}[a_2F^{LL}_{Aa_0}+C_2 M^{LL}_{Aa_0}+a_1F^{LL}_{Th}+C_1 M^{LL}_{Th}]\nonumber\\
&-&V_{tb}^*V_{td}[(a_3+a_9-a_5-a_7)F^{LL}_{Aa_0}+(C_4+C_{10})M^{LL}_{Aa_0}\nonumber\\
&+&(C_6+C_8)M^{SP}_{Aa_0}+(a_4+a_{10})F^{LL}_{Th}+(a_6+a_8)F^{SP}_{Th}\nonumber\\
&+&(C_3+C_9)M^{LL}_{Th}+(C_5+C_7)M^{LR}_{Th}+(\frac{4}{3}(C_3+C_4-\frac{C_9}{2}-\frac{C_{10}}{2})\nonumber\\
&-&a_5+\frac{a_7}{2})F^{LL}_{Ah}+(a_6-\frac{a_8}{2})F^{SP}_{Ah}+(C_3+C_4-\frac{C_9}{2}-\frac{C_{10}}{2})M^{LL}_{Ah}\nonumber\\
&+&(C_5-\frac{C_7}{2})M^{LR}_{Ah}+(C_6-\frac{C_8}{2})M^{SP}_{Ah}]\big\} \;,
\label{amp5}
\eeq
 \beq
 {\cal A}(B^0 \to a_0^0\pi^0) &=& \frac{G_F} {2\sqrt{2}}
 \big\{V_{ub}^*V_{ud}[a_2(F^{LL}_{Aa_0}+F^{LL}_{Ah}-F^{LL}_{Ta_0}-F^{LL}_{Th})+C_2(M^{LL}_{Aa_0}+M^{LL}_{Ah}\nonumber\\
&-&M^{LL}_{Ta_0}-M^{LL}_{Th})]-V_{tb}^*V_{td}[(a_4-\frac{5 C_9}{3}-C_{10}+\frac{3 a_7}{2})F^{LL}_{Ta_0}\nonumber\\
&+&(a_6-\frac{a_8}{2})(F^{SP}_{Ta_0}+F^{SP}_{Aa_0}+F^{SP}_{Th}+F^{SP}_{Ah})+(C_3-\frac{C_9+3 C_{10}}{2})(M^{LL}_{Ta_0}+M^{LL}_{Th})\nonumber\\
&+&(C_5-\frac{C_7}{2})(M^{LR}_{Ta_0}+M^{LR}_{Aa_0}+M^{LR}_{Th}+M^{LR}_{Ah})-\frac{3 C_8}{2}(M^{SP}_{Ta_0}+M^{SP}_{Th})\nonumber\\
&+&(\frac{7 C_3+5 C_4+C_9-C_{10}}{3}-2 a_5-\frac{a_7}{2})(F^{LL}_{Aa_0}+F^{LL}_{Ah})\nonumber\\
&+&(C_3+2 C_4-\frac{C_9- C_{10}}{2})(M^{LL}_{Aa_0}+M^{LL}_{Ah})+(2 C_6+\frac{C_8}{2})(M^{SP}_{Aa_0}+M^{SP}_{Ah})\nonumber\\
&+&(a_4-\frac{5 C_9}{3}-C_{10}-\frac{3 a_7}{2})F^{LL}_{Th}] \;,
\label{amp6}\\
 {\cal A}(B^0 \to a_0^-\pi^+) &=& \frac{G_F} {\sqrt{2}}
 \big\{V_{ub}^*V_{ud}[a_1 F^{LL}_{Ta_0}+a_2 F^{LL}_{Ah}+C_1 M^{LL}_{Ta_0}+C_2 M^{LL}_{Ah}]-V_{tb}^*V_{td}[(a_4\nonumber\\
&+&a_{10})F^{LL}_{Ta_0}+(a_6+a_8)F^{SP}_{Ta_0}+(C_3+C_9)M^{LL}_{Ta_0}+(C_5+C_7)M^{LR}_{Ta_0}\nonumber\\
&+&(\frac{4}{3}(C_3+C_4-\frac{C_9+C_{10}}{2})-a_5+\frac{a_7}{2})F^{LL}_{Aa_0}+(a_6-\frac{a_8}{2})F^{SP}_{Aa_0}\nonumber\\
&+&(C_3+C_4-\frac{C_9+C_{10}}{2})M^{LL}_{Aa_0}+(C_5-\frac{C_7}{2})M^{LR}_{Aa_0}+(C_6-\frac{C_8}{2})M^{SP}_{Aa_0}\nonumber\\
&+&(a_3+a_9-a_5-a_7)F^{LL}_{Ah}+(C_4+C_{10})M^{LL}_{Ah}+(C_6+C_8)M^{SP}_{Ah}]\big\} \;,
\label{amp7}\\
 {\cal A}(B^0 \to a_0^0K^0) &=& \frac{G_F} {2}
 \big\{V_{ub}^*V_{us}[a_2 F^{LL}_{Th}+C_2 M^{LL}_{Th}]-V_{tb}^*V_{ts}[-(a_4-\frac{a_{10}}{2})(F^{LL}_{Ta_0}+F^{LL}_{Aa_0})\nonumber\\
&-&(a_6-\frac{a_8}{2})(F^{SP}_{Ta_0}+F^{SP}_{Aa_0})-(C_3-\frac{C_9}{2})(M^{LL}_{Ta_0}+M^{LL}_{Aa_0})\nonumber\\
&-&(C_5-\frac{C_7}{2})(M^{LR}_{Ta_0}+M^{LR}_{Aa_0})+\frac{3}{2}(a_7+a_9)F^{LL}_{Th}+\frac{3 C_{10}}{2}M^{LL}_{Th}\nonumber\\
&+&\frac{3 C_8}{2}M^{SP}_{Th}]\big\} \;,
\label{amp8}\\
 {\cal A}(B^0 \to a_0^-K^+) &=& \frac{G_F} {\sqrt{2}}
 \big\{V_{ub}^*V_{us}[a_1 F^{LL}_{Ta_0}+C_1 M^{LL}_{Ta_0}]-V_{tb}^*V_{ts}[(a_4+a_{10})F^{LL}_{Ta_0}\nonumber\\
 &+&(a_6+a_8)F^{SP}_{Ta_0}+(C_3+C_9)M^{LL}_{Ta_0}+(C_5+C_7)M^{LR}_{Ta_0}\nonumber\\
&+&(a_4-\frac{a_{10}}{2})F^{LL}_{Aa_0}+(a_6-\frac{a_8}{2})F^{SP}_{Aa_0}+(C_3-\frac{C_9}{2})M^{LL}_{Aa_0}\nonumber\\
&+&(C_5-\frac{C_7}{2})M^{LR}_{Aa_0}]\big\} \;,
\label{amp9}\\
 {\cal A}(B_s^0 \to a_0^+\pi^-) &=& \frac{G_F} {\sqrt{2}}
 \big\{V_{ub}^*V_{us}[a_2 F^{LL}_{Aa_0}+C_2 M^{LL}_{Aa_0}]-V_{tb}^*V_{ts}[(a_3+a_9-a_5-a_7)F^{LL}_{Aa_0}\nonumber\\
&+&(C_4+C_{10})M^{LL}_{Aa_0}+(C_6+C_8)M^{SP}_{Aa_0}+(a_3-\frac{a_9}{2}-a_5+\frac{a_7}{2})F^{LL}_{Ah}\nonumber\\
&+&(C_4-\frac{C_{10}}{2})M^{LL}_{Ah}+(C_6-\frac{C_8}{2})M^{SP}_{Ah}]\big\} \;,
\label{amp10}
\eeq
 \beq
 {\cal A}(B_s^0 \to a_0^0\pi^0) &=& \frac{G_F} {2\sqrt{2}}
 \big\{V_{ub}^*V_{us}[a_2(F^{LL}_{Aa_0}+F^{LL}_{Ah})+C_2(M^{LL}_{Aa_0}+M^{LL}_{Ah})]\nonumber\\
&-&V_{tb}^*V_{ts}[(2 a_3+\frac{a_9}{2}-2 a_5-\frac{a_7}{2})(F^{LL}_{Aa_0}+F^{LL}_{Ah})\nonumber\\
&+&(2C_4+\frac{C_{10}}{2})(M^{LL}_{Aa_0}+M^{LL}_{Ah})+(2C_6+\frac{C_8}{2})(M^{SP}_{Aa_0}+M^{SP}_{Ah})]\big\} \;,
\label{amp11}\\
 {\cal A}(B_s^0 \to a_0^-\pi^+) &=& \frac{G_F} {\sqrt{2}}
 \big\{V_{ub}^*V_{us}[a_2 F^{LL}_{Ah}+C_2 M^{LL}_{Ah}]-V_{tb}^*V_{ts}[(a_3-\frac{a_9}{2}-a_5+\frac{a_7}{2})F^{LL}_{Aa_0}\nonumber\\
 &+&(C_4-\frac{C_{10}}{2})M^{LL}_{Aa_0}+(C_6-\frac{C_8}{2})M^{SP}_{Aa_0}+(a_3+a_9-a_5-a_7)F^{LL}_{Ah}\nonumber\\
 &+&(C_4+C_{10})M^{LL}_{Ah}+(C_6+C_8)M^{SP}_{Ah}]\big\} \;,
 \label{amp12}\\
 {\cal A}(B_s^0 \to a_0^+K^-) &=&  \frac{G_F} {\sqrt{2}} \big\{V_{ub}^*V_{ud}[a_1 F^{LL}_{Th}+C_1 M^{LL}_{Th}]-V_{tb}^*V_{td}[(a_4+a_{10})F^{LL}_{Th}\nonumber\\
&+&(a_6+a_8)F^{SP}_{Th}+(C_3+C_9)M^{LL}_{Th}+(C_5+C_7)M^{LR}_{Th}+(a_4-\frac{a_{10}}{2})F^{LL}_{Ah}\nonumber\\
&+&(a_6-\frac{a_8}{2})F^{SP}_{Ah}+(C_3-\frac{C_9}{2})M^{LL}_{Ah}+(C_5-\frac{C_7}{2})M^{LR}_{Ah}]\big\} \;,
\label{amp13}\\
 {\cal A}(B_s^0 \to a_0^0 \bar K^0) &=& \frac{G_F} {2}
 \big\{V_{ub}^*V_{ud}[a_2 F^{LL}_{Th}+C_2 M^{LL}_{Th}]-V_{tb}^*V_{td}[(\frac{5 C_9}{3}+C_{10}+\frac{3a_7}{2}-a_4)F^{LL}_{Th}\nonumber\\
&-&(a_6-\frac{a_8}{2})(F^{SP}_{Th}+F^{SP}_{Ah}) +(\frac{C_9}{2}+\frac{3 C_{10}}{2}-C_3)M^{LL}_{Th}-(C_5-\frac{C_7}{2})(M^{LR}_{Th}+M^{LR}_{Ah})\nonumber\\
&+&\frac{3 C_8}{2}M^{SP}_{Th}-(a_4-\frac{a_{10}}{2})F^{LL}_{Ah}-(C_3-\frac{C_9}{2})M^{LL}_{Ah}]\big\} \;,
\label{amp14}
\eeq
In these expressions, $G_F$ is the fermi coupling constant, $V$'s are the CKM matrix elements,
the combined Wilson coefficients $a_i$ are defined as
\beq
&& a_1=C_2+\frac{C_1}{3}, \quad\quad a_2= C_1+\frac{C_2}{3},\non
&& a_i= C_i+\frac{C_{i+1}}{3} \quad\quad {\rm with} \, i=3-10 \,.
 \eeq
The factorizable and nonfactorizable amplitudes, saying $F$ and $M$ respectively, can be found in Refs. \cite{Wang:2020saq}.



\end{document}